# Cheaper is not enough: fleet inertia, network confidence and the limits of diesel pricing in Australia's zero-emission truck transition

Elnaz Irannezhad[1], Seyed Hossein Hosseini[1,2], David A. Hensher[3]

[1] Research Centre for Integrated Transport Innovations (rCITI), School of Civil and Environmental Engineering, University of New South Wales (UNSW), Sydney, NSW, Australia

[2] Sydney School of Public Health, Faculty of Medicine and Health, The University of Sydney, Sydney, NSW, Australia

[3] Institute of Transport and Logistics Studies (ITLS), Business School, The University of Sydney, Sydney, NSW, Australia

## Abstract

What would it take to accelerate the adoption of zero-emission trucks? Even if battery electric trucks become cheaper to own over their lifetime than diesel trucks, that alone is not enough. This study develops a system dynamics model of the rigid and articulated truck fleet, calibrated to the Australian fleets. Under current policy regimes, more than a third of the fleet still runs on diesel in 2050. Battery and fuel cell trucks become as attractive as diesel trucks only in 2031–2032, held back by slow fleet turnover and by how long operators take to trust the charging network, the most influential assumption tested in this study. Taxing diesel barely works at politically feasible levels. Charging trucks the full fuel excise lifts the zero-emission share by one percentage point; adding ten points would require a road user charge of about 5 AUD per litre, fifteen times the 2025–26 rate. A 70 AUD per tonne carbon price and a purchase subsidy covering 60% of the price premium do little better. Per percentage change in cost, operating-cost support is five to six times as effective as purchase support, yet the best single instrument adds only two points. Only a combined push–pull package, which still raises fuel-tax revenue, shifts the market materially. Policy should stop relying on diesel prices and instead build operators' confidence in the network, cut operating costs and recycle road-charge revenue into infrastructure.

**Keywords:** Road Freight Decarbonisation; System Dynamics; Heavy Vehicle; Total Cost of Ownership; Charging Infrastructure; Battery Truck; Hydrogen Fuel Cell Truck; Australia

# INTRODUCTION

Globally, there has been noticeable growth in the adoption of battery electric vehicles in the passenger sector and buses. However, the broader decarbonisation of the transport sector, encompassing heavy freight vehicles, remains slow and fragmented. The Heavy vehicle (HV) sector faces well-known barriers, including limited commercial availability of electric trucks, battery size and distance range, and insufficient charging and refuelling infrastructure (Gerber Machado et al., 2020). Particularly, in some countries such as Australia, Canada and the United States (US), the freight industry favours high-productivity multi-trailer HVs designed for larger loads, posing a further challenge for electrification. Although larger HVs are generally more energy-efficient on a tonne-kilometre basis, their transition to zero-emission technologies remains constrained by the limited commercial availability of battery-electric and hydrogen fuel cell models. This challenge is compounded by the concentration of major Original Equipment Manufacturers (OEMs) in Europe and China, where vehicle designs, regulatory priorities, and freight operating conditions differ, limiting the near-term availability of suitable models for the other markets such as Australia, Canada, and US.

As the model availability and total cost of ownership (TCO) of battery electric rigid trucks and light commercial vehicles are getting close to internal combustion engine (ICE) trucks, the growth of zero-emission heavy vehicle (ZEHV) sales is noticeable. In Europe, the zero-emission share of new vehicles of 3.5–12 tonnes, a category that combines medium trucks and heavy vans, doubled from 10.4% in 2024 to 21% in 2025. Comparatively, the sales of heavy-duty trucks were much less (~1.5%) (International Council on Clean Transportation, 2026b). The long-haul HV prime movers require massive horsepower and torque, hence bigger battery sizes and as a result would reduce the payload and the desired load productivity. Transport operators are also often concerned about reducing TCO to improve their competitiveness. Given that the freight transport cost is a significant component of consumer goods price, any small increase in total transport costs may result in inflation. The literature suggests that the TCO of battery electric trucks (BETs) and hydrogen fuel cell trucks (FCTs) was significantly higher than that of ICE trucks in UK in 2022 (Rout et al., 2022). Given the depot charging costs and high purchase price of ZEHVs, the sector's fleet transition would require a sound economic justification or regulatory reforms.

Compounding these challenges is a siloed approach to assessing the effectiveness of various policies. Studies on decarbonising the HV sector suggest that a linear carbon tax increase to $150/t CO2e by 2032 could reduce emissions by 43% in 2060 (Keller et al., 2019), while transitioning to a fleet of BET alone may fall short of the required 30% emissions reduction targets without increased renewable energy sources (Sen et al., 2019). On the regulatory side, the EU's 15% $CO_2$ reduction target for new HVs, which applies from the 2025 reporting period (1 July 2025 to 30 June 2026) under Regulation (EU) 2019/1242 as amended by Regulation (EU) 2024/1610 (European Parliament, 2024), coincided with a threefold increase in the sales of ZEHVs over 12 tonnes in Europe compared with 2024. In the United States, however, the ZEHV market share shrank by 3.6% compared to the year before, due to removing some of the decarbonisation targets (International Council on Clean Transportation, 2026c). Undoubtedly, a range of policies can be implemented to accelerate the transition to low-emission freight transport and reduce tailpipe emissions. These include increasing fuel excise taxes or diesel prices, introducing fuel efficiency standards, and providing subsidies for emissions reduction technologies, such as after-treatment systems for existing ICE trucks.

Current freight decarbonisation policies often rely on simplified assumptions about the dynamic and interdependent relationship between demand-side change and supply-side constraints. In particular, existing TCO and adoption models commonly assume relatively linear relationships between fuel prices, vehicle costs, and demand, while giving limited attention to feedback loops such as fleet turnover, infrastructure rollout, and scale effects. As a result, they may not adequately capture the “chicken-and-egg” problem that characterises the transition to ZEHVs, where vehicle uptake depends on infrastructure availability and infrastructure investment depends on expected uptake.

Australia’s HV charging and refuelling network is still at an early stage. Most battery electric trucks in service are charged at operators’ depots, often supported by on-site solar generation and storage; for example, Centurion’s two off-grid depots at Perth Airport, co-funded by the Australian Renewable Energy Agency (ARENA), now charge a fleet of 30 electric trucks (CleanTechnica, 2026). The first open-access truck charging network is being built in Melbourne, where ARENA is co-funding three shared hubs with 24 high-power bays at Laverton North, Somerton and Dandenong South, due to open between late 2026 and mid-2027 as the first stage of a network planned along the Hume, Pacific and Newell highways (The Driven, 2026, Global Fleet, 2026). ARENA estimates that up to 165 HV charging hubs may be needed nationally (Global Fleet, 2026). Battery swapping is also being trialled, with nine swap stations operating for converted electric prime movers (CleanTechnica, 2026). Hydrogen refuelling for HVs is limited to a few sites, such as Port Kembla in New South Wales and Geelong in Victoria, while the jointly funded Hume Hydrogen Highway envisages at least four stations between Sydney and Melbourne (Coregas, 2023, Victorian Government, 2025). In such a sparse network, mobile or relocatable charging units, for example battery trailers or containerised chargers that can be positioned at freight hubs, ports and corridor rest areas where demand is emerging, could bridge gaps while demand remains uncertain and reduce the risk of stranded fixed investment.

The urgency of this research is underscored by the Strait of Hormuz crisis in the context of the Iran–US conflict. Attacks on shipping and the effective closure of the strait pushed the price of Brent crude oil from about 65 USD per barrel in January 2026 to above 100 USD per barrel in mid-March 2026 and again in early October 2026, an increase of more than 50%, with diesel prices rising sharply across global markets (Australian Institute of Petroleum, 2026). These developments have immediate and widespread impacts on freight transport costs, inflation and supply chains, highlighting the systemic risks of continued reliance on imported fossil fuels. Freight decarbonisation is therefore not only a long-term climate objective but also a strategy for enhancing energy security and economic resilience.

The most recent study by Raoofi et al. (2025a) presented a system dynamics (SD) model and examined how policy instruments affect energy efficiency and BET adoption in Sweden. However, that study did not differentiate between HV sizes. This is an important limitation because the feasibility of electrification differs substantially across rigid trucks (RTs) and articulated trucks (ATs). In the Australian, Canadian or American context, these differences are especially important because freight movements often rely on high-productivity vehicle combinations and long-distance operations, which create stronger constraints for BET and FCT adoption than those typically observed in Europe.

Hensher and Wei (2024) developed a comprehensive decision support system (DSS) for the fleet transition in Australia to quantify the costs to industry and total emissions and air pollutants under various decarbonisation strategies of converting Euro V trucks to either the low-emission Euro VI trucks or ZEHVs from 2030 to 2040. The total emission and costs are calculated based on the user input values for average tonne and kilometres travelled for three HV sizes (i.e. rigid, articulated and B-double articulated trucks), as well as other known parameters, including energy usage, energy costs, emission and air pollutants of each energy type, and number of trucks of different classes. Another study by Hosseini et al. (2025) developed an SD model for HV energy consumption in Australia, categorised for LCVs, RTs and ATs. Notably, these two studies presented the fleet transition for various truck sizes, particularly the breakdown of rigid trucks versus articulated trucks under different decarbonisation pathways.

This current study extends the research by integrating detailed ownership cost dynamics and charging and refuelling infrastructure feedback loops and develops an SD-based decision-support framework for the Australian HV sector that integrates truck segment choice, fleet turnover, energy price dynamics, charging infrastructure rollout, and policy interventions within a single modelling structure. The model examines two HV segments (RTs and ATs) and evaluates the combined effects of push and pull policies under alternative scenarios. This study also extends the analysis beyond BET by including hydrogen FCTs, which may be more relevant for high-productivity and long-haul applications. The aim is to facilitate the testing of a bundle of various policy measures, with a focus on fiscal measures aimed at overcoming transition barriers.

More specifically, this research attempts to address these research questions: (i) How do fleet turnover, energy prices, and charging infrastructure availability interact to shape the adoption of ZEHVs in countries such as Australia? (ii) How do transition pathways differ between RTs and ATs? (iii) Which combinations of push and pull policies are most effective for different HV segments? (iv) Which vehicle size is more elastic with respect to the decarbonisation policies?

Drawing on the system dynamics and TCO literature reviewed in Section 2, we formulate six testable hypotheses that structure the analysis:

**H1 (fleet inertia).** Under current policy settings, slow fleet turnover and the gradual build-up of charging and refuelling infrastructure keep zero-emission truck uptake well below the pace required for a zero-emission rigid and articulated fleet by 2050.

**H2 (durability of price signals).** A persistent, two-sided widening of the price gap between diesel and electricity or hydrogen produces a larger long-run shift in market share than a front-loaded fossil-fuel price shock of comparable magnitude.

**H3 (operating versus purchase costs).** Instruments acting on operating costs (OPEX) shift adoption more than instruments acting on purchase costs (CAPEX), because OPEX dominate the TCO (Noll et al., 2022, Rout et al., 2022).

**H4 (push versus pull).** Push instruments that raise the OPEX of diesel trucks, such as road user charges and carbon pricing, accelerate adoption more than pull instruments that lower the OPEX of BET/FCT.

**H5 (segment heterogeneity).** Because long-haul articulated operations face more demanding range, payload and infrastructure requirements, ATs are less responsive to decarbonisation policies than RTs.

**H6 (infrastructure feedback).** In high-uptake scenarios, charging and refuelling capacity becomes a binding balancing constraint that dampens further adoption.

The objective of this paper is therefore to quantify how far, how fast and with what fiscal consequences alternative energy-price conditions and push–pull policy packages can shift Australia's rigid and articulated truck fleet towards battery electric and fuel cell technologies, and to identify which policy lever and which vehicle segment offer the greatest leverage. The paper makes four contributions. First, it embeds a fleet turnover model, a discounted TCO formulation and a charging and refuelling infrastructure sector driven by demand and station profitability within a single SD model calibrated to Australian data, distinguishing RTs from ATs and BETs from FCTs. Second, it separates the OPEX and CAPEX channels, and the push and pull sides, of decarbonisation policy, reporting dynamic elasticities and the timing of relative-attractiveness parity alongside market shares. Third, it quantifies the fiscal feedback of the transition. Push measures raise fuel-tax revenue while pull measures, and the transition itself, erode it, which shows why push and pull instruments need to be combined. Fourth, it provides quantitative evidence for a market characterised by high-productivity articulated combinations, no mandatory HV $CO_2$ standards and limited zero-emission truck incentives, a setting that Australia shares with other markets such as Canada and the United States.

The paper organisation is as follows. Section 2 reviews the literature and Section 3 describes the Australian case study. Section 4 presents the methods, data and model validation. Section 5 presents the results and sensitivity analysis, and Section 6 assesses the hypotheses and discusses the implications and concludes.

# LITERATURE REVIEW

## System Dynamics Studies

System dynamics (SD) modelling, pioneered by Forrester (1961), is a simulation-based methodology that captures the dynamic, non-linear behaviour of complex socio-technical systems through stocks, flows, reinforcing feedback loops, and delays. It is particularly suited to analysing technology diffusion and policy impacts in transportation, where adoption is influenced by interdependent factors such as costs, infrastructure, regulations, awareness, and stakeholder behaviours. In the context of freight decarbonisation, SD models have been employed to simulate fleet renewal, market penetration of alternative powertrains, and the ripple effects on energy demand, emissions, and infrastructure (Ghisolfi et al., 2022). SD models illuminate the feedback mechanisms, e.g., reinforcing loops from cost reductions via scale and balancing loops from infrastructure bottlenecks, that determine adoption trajectories.

Ghisolfi et al. (2022) undertook a systematic review of SD models in freight transport decarbonisation, and identified 13 models focused on switching to lower-carbon energy sources, including electrification and alternative fuels. These models highlight the role of subsidies, infrastructure investment, and technology learning curves in driving transitions but reveal persistent gaps, namely limited integration of multiple strategies (e.g., electrification alongside mode shift), insufficient attention to time lags and endogenous policy dynamics, and a lack of transparency in model equations and validation. This review notes that while SD excels at subsystem interactions (e.g., fleet-infrastructure), integrated models for competing zero-emission technologies remain underdeveloped, often siloed by region or fuel type.

Earlier SD studies on alternative powertrains HVs (e.g. (Seitz and Terzidis, 2014)), emphasised the interplay of costs, research and development (R&D) and regulations in market penetration, setting the foundation for more recent work on ZEHVs. Focused on Sweden, Raoofi et al. (2024) presented causal loop diagrams (CLDs) to map system-level impacts of road freight electrification, identifying two adoption phases: an early infrastructure-constrained phase and a later efficiency-driven phase. Raoofi et al. (2025b) extended this by linking BET fleets to electricity demand, revealing feedback where high adoption raises prices, potentially slowing diffusion unless supply-side investments occur. The model integrated stocks of BET fleets and charging infrastructure, with flows driven by technology maturity, operator awareness, TCO, and subsidies. Key reinforcing loops included falling vehicle costs from R&D investments and network effects from charging expansion; balancing loops arise from grid constraints and diesel competition. Policy simulations revealed that charging station subsidies yield the highest adoption gains (e.g., substantial fleet growth by 2060), while R&D subsidies for technology maturity prove most cost-efficient under budget constraints. The study underscored endogenous policy dynamics. Subsidies accelerate initial uptake but risk diminishing returns if not paired with infrastructure scaling. Similarly, Raoofi et al. (2025a) analysed efficiency gains from electrification, showing reduced logistics costs but potential station underutilization without optimised deployment.

Beyond Europe, Abaddi (2025) applied the SD approach to BET adoption in Jordan, incorporating local logistical operations, cost structures, and policy incentives. The model simulated fleet transitions under scenarios of subsidies and infrastructure rollout, demonstrating that targeted incentives can achieve 20–30% market share by 2040, contingent on overcoming high upfront costs and grid limitations.

The SD studies often treat hydrogen as peripheral or absent. A notable exception is a study by Kotlars et al. (2025), which quantifies competition via multi-factor attractiveness (acquisition, fuel, maintenance) weighted by operator preferences, through a tailored SD model for small economies. This study uses a generalised Bass diffusion structure augmented with subsystems for fleet renewal, infrastructure, costs, and attractiveness, reinforcing innovation (cost learning from production and industrial hydrogen demand) and balancing constraints (high station rollout costs in limited markets). Comparing FCTs to BETs, results showed BETs dominate under market-led scenarios due to lower entry barriers (e.g., 350,000 EUR acquisition vs. 450,000 EUR for FCTs), but FCTs achieve parity with sustained subsidies (up to 120,000 EUR/truck) and cross-sector demand (e.g., "hydrogen valleys" linking transport and industry). Sensitivity analyses highlighted leverage points. First, early infrastructure subsidies and learning rates (25% cost reduction per demand doubling) are critical, with hydrogen excelling in cold-chain logistics. Another study in China by Teng et al. (2024) employed a modified Bass model with endogenous technology learning–aligned with SD principles–to evaluate policy mixes for heavy-duty FCTs across ATs, dump trucks, and RTs. Simulating 113 scenarios to 2060, they found sales bans and hydrogen subsidies indispensable, boosting ownership by up to 35.7% and cutting social costs by 35% in optimal mixes. While not purely SD, the model's diffusion mechanics echo SD feedback, positioning hydrogen as complementary to BETs for heavy-duty applications. Hosseini et al. (2025) also investigated the HV fleet transition in Australia, but overlooked the charging infrastructure feedback loop, and the different truck sizes of BETs and FCTs.

These studies collectively affirm SD's strength in capturing long-term, path-dependent behaviours, such as the "chicken-and-egg" problem of vehicles and chargers. However, existing SD studies in freight decarbonisation often incorporate cost-related adoption mechanisms but provide limited integration of endogenous TCO dynamics, infrastructure rollout, and fleet turnover feedbacks (Ghisolfi et al., 2022). The next subsection presents a literature review of TCO studies related to the HV sector.

### Total Cost of Ownership Studies

TCO studies of BETs and FCTs, reviewed in Appendix A, consistently find that OPEX account for about three-quarters of the TCO of HVs because of their high annual mileage, so that the competitiveness of ZEHVs depends mainly on energy prices, road charges and tolls rather than on purchase prices (Noll et al., 2022). Purchase grants reduce the TCO of ZEHVs by only about 3–4%, whereas fossil-fuel taxes raise the TCO of diesel trucks substantially (Rout et al., 2022). Most of these studies, however, compare fixed vehicle configurations at given points in time and do not represent fleet turnover, infrastructure feedbacks or the response of purchase decisions to changing costs, which motivates the dynamic approach of this study.

### Literature on Infrastructure, Energy Consumption, and Policy Interventions

Several studies have assessed the infrastructure investment requirements for BET and FCTs. Breuer et al. (2021) investigated various ZEHV powertrains while considering the necessary investment and costs for required infrastructure. Exploring the technical feasibility of medium and heavy-duty FCTs and BETs, Forrest et al. (2020) assessed performance and infrastructure availability, using California as a reference for determining electric load demand and its ability to meet state demand. Kotze et al. (2021) investigated the necessary investments to decarbonise the HV sector using green hydrogen in New Zealand, employing a customised methodology combining SD modelling and hydrogen transition modelling.

Several studies have quantified the energy consumption and emissions of HVs, comparing BETs and FCTs with ICE trucks. Ally et al. (2015) highlighted that energy use and emissions from HVs are growing at a faster rate compared to light-duty vehicles and are projected to surpass them by 2032. Talebian et al. (2018) examined the potential for ZEHVs to achieve a 64% reduction in greenhouse gas emissions from freight road transport by 2040 in British Columbia. Sun et al. (2021) focused on evaluating the fuel economy, emissions and range of FCTs by conducting tests on one trial truck under different driving cycles. Another study by Guandalini and Campanari (2018) conducted a similar well-to-wheel analysis, comparing BETs and FCTs, and concluded that both vehicle types are competitive in terms of energy consumption and efficiency, with outcomes varying based on specific scenarios. Booto et al. (2021) investigated the environmental impacts of ICE, BET, and FCTs over their lifetimes, considering the well-to-wheel approach throughout the entire value chain in Norway. Haugen et al. (2021) investigated diverse energy pathways for vehicles and assessed their impact on energy efficiency and CO2 reduction. Their findings highlighted that BETs achieved the highest energy efficiency in terms of km/1000 MJ and exhibited lower CO2 emissions compared to FCTs.

Watabe et al. (2020) examined the impact on GHG emissions when vehicles powered by fossil fuels are progressively banned in Japan starting from 2035. Their evaluation of different scenarios, including renewable energy use, refuelling station infrastructure development, and hydrogen production methods, indicated that implementing a vehicle ban for internal combustion vehicles could result in life-cycle GHG emission reductions of up to 438 Mt by 2060 compared to 2017 levels. Mac Kinnon et al. (2016) focused on the air-quality impacts of adopting FCT in light and heavy-duty vehicles in California, and found significant improvements in air quality, particularly a substantial reduction in fine particulate matter (PM2.5). The study highlighted that decarbonising HVs would have a greater impact on air pollution reduction compared to light vehicles, providing insights for policymakers to implement impactful policies.

Various country-specific studies discussed the policy implications for the acceleration of BET/FCT fleet transition, such as fuel efficiency improvements in Iceland (Shafiei et al., 2014), carbon tax in Germany (Cerniauskas et al., 2019), policy mandates and incentives in California, USA (Burke and Miller, 2020), and the “platform licenses” in China that included priority dispatch and commission discounts for drivers (Yan et al., 2022). For a comprehensive review of current European legislation and standardisation for ZEHVs, please refer to the review paper by Borbujo et al. (2021). This review paper highlights key areas where standards should be updated as technology evolves, such as "higher capacity and flow charging rates”. Burke and Miller (2020) and Burke and Sinha (2020) evaluated potential policy mandates and incentives for supporting the implementation of ZEHV in California. These studies specifically examined the effects of policies, projected the incentives needed for various market share scenarios, and recommended the simultaneous application of mandates and incentives to expedite adoption.

# AUSTRALIA'S HV DECARBONISATION POLICY AND GLOBAL COMPARISON

The reviewed literature focuses largely on European jurisdictions whose policy settings strongly favour BETs and FCTs. Table 1 compares the main HV decarbonisation instruments in three of the leading markets, Norway, Sweden and Germany, with those in Australia. The three European countries combine several mutually reinforcing instruments: binding fleet-wide $CO_2$ standards for new HVs (through EU regulation in Sweden and Germany) or an explicit phase-out goal (Norway); road charges that exempt or discount ZEHVs, such as Germany's distance-based truck toll, which has included a $CO_2$ component of 200 Euro per tonne since December 2023 while exempting ZEHVs until mid-2031; explicit carbon pricing of transport fuels; purchase support covering a substantial share of the cost premium of ZEHVs (Norway and Sweden); and public funding for HV charging. The low-carbon, hydro- and nuclear-dominated electricity systems of Norway and Sweden further lower the OPEX and life-cycle emissions of BETs. The effects of these policy packages are now visible in the market share. ZEHVs accounted for about one in six new heavy lorries in Norway in mid-2026, close to one in ten new heavy trucks in Sweden and 4.6% in Germany, compared with a very small share in Australia.

Australia's settings differ on almost every dimension. There are no $CO_2$ or fuel-efficiency standards for HVs. The newly introduced Vehicle Efficiency Standard in 2025 applies only to light vehicles, and ADR 80/04, which is based on Euro VI and has been mandatory for new HV models since November 2024 and for all new HVs supplied since November 2025, regulates only noxious pollutants. Transport fuels are not subject to a carbon price, and HVs on public roads recover all of the diesel excise except the RUC through fuel tax credits, a charge that was itself temporarily suspended in 2026 (see below). Support for alternative fuel trucks has so far been project-based rather than economy-wide. The Clean Energy Finance Corporation (CEFC) has committed 70 million AUD with Volvo Group Australia to provide discounted finance and residual-value support for locally built electric trucks, the first of which left the Brisbane production line in July 2026 (Prime Minister of Australia, 2026), and the Australian Renewable Energy Agency (ARENA) has co-funded electric truck fleet deployments. Mass and width concessions for ZEHVs are emerging, but they vary between states and are largely permit-based. These initiatives are valuable for building operating experience and a local supply chain, but, unlike the European packages, they do not alter the relative OPEX faced by most operators, which our results identify as the most powerful lever for adoption.

Australia's only economy-wide carbon price operated from July 2012 to July 2014, at 23 AUD per tonne of $CO_2$-e in its first year, and was repealed before the planned transition to an emissions trading scheme (Climate Council, 2015); transport fuels have not been subject to a carbon price since. The fuel excise tax currently represents one of the primary policy instruments for internalising the negative externalities of fossil fuel consumption. Fuel excise tax, imposed on the sale of fuel, is a substantial income source for the Australian government in funding infrastructure projects, development and maintenance. Historically associated with road development, the majority of fuel excise is now collected as general income, financing diverse public services such as healthcare, education, and the military. The excise does not provide a specific allocation for these sectors, as it contributes to the total government budget.

**Table 1 | HV decarbonisation policy settings in Norway, Sweden, Germany and Australia (as of October 2026).**

| Instrument | Norway | Sweden | Germany | Australia |
|---|---|---|---|---|
| **$CO_2$ standards or targets for new HVs** | Parliamentary goal (2023): all new trucks zero-emission or biogas-fuelled from 2030 | EU HV $CO_2$ standards: −15% (2025–29), −45% (2030–34), −65% (2035–39), −90% (from 2040) vs 2019 | EU HV $CO_2$ standards (as for Sweden) | None; ADR 80/04 (Euro VI) covers noxious pollutants only; NVES covers light vehicles only |
| **Road charges** | Reduced, often zero, toll rates for zero-emission HVs | Time-based road charge for trucks >12 t, $CO_2$-differentiated since 2025 | Distance-based truck toll with 200Euro/t $CO_2$ component (since Dec 2023); ZEHVs exempt until 30 June 2031 | RUC of 32.4 c/L collected through diesel fuel tax credits; suspended Apr–Jun 2026 and set at 16.4 c/L from 1 Jul and restored to 32.4 c/L from 3 Aug 2026 |
| **Carbon pricing and fuel taxation of diesel** | $CO_2$ tax NOK 4.42/L plus road-usage tax NOK 2.28/L (2026; temporarily reduced in mid-2026) | Energy and $CO_2$ taxes plus a biofuel reduction obligation; temporary tax cuts in 2026 | National $CO_2$ price of 55–65Euro/t (2026), to be replaced by EU ETS2 from 2028 | No carbon price (repealed in 2014); on-road HVs recover all excise except the RUC |
| **Purchase or finance support** | Enova competitive grants covering up to 60% of the extra cost over a diesel truck | Up to 25% of the purchase price of electric or gas trucks (periodic calls) | Purchase subsidy (80% of extra cost) closed in 2024; no national truck purchase subsidy in 2026 | No general subsidy; 70 million AUD discounted finance and residual-value support (Volvo); ARENA fleet-trial grants |
| **Charging and refuelling infrastructure** | Enova support for HV charging; National Transport Plan 2025–2036 prioritises charging at rest areas | Truck charging pools every 60 km on the TEN-T core network by 2030 | EU AFIR plus 1 billion Euro (2026–2029) for depot and non-public HV charging | No national HV charging plan; project-based CEFC and ARENA co-funding |
| **Mass and dimension concessions** | Higher gross weight limits for zero-emission and biogas trucks | EU rules allow up to 2 t extra for zero-emission combinations | EU rules allow up to 2 t extra for zero-emission combinations | State-based, mostly permit-based steer-axle and width concessions (e.g. NSW up to 8.0 t single steer) |
| **Zero-emission share of new heavy trucks** | 10.3% (2025); about one in six (mid-2026) | 9.7% (2026, >12 t); about 10% (Jan–Aug 2026) | 1.8% (2025); 4.6% (2026), >12 t | Very low (well below 1%); no official powertrain breakdown |

*Sources: Norway (Statens, 2026, Enova, 2024, Skatteetaten, 2026); Sweden (Energimyndigheten, 2026, Transportstyrelsen, 2025); Germany (Bundesministerium für Verkehr, 2025, Bundesamt für Logistik und Mobilität, 2025, Deutsche Emissionshandelsstelle, 2026); EU (European Parliament, 2024, European Parliament, 2023); Australia (Australian Taxation Office, 2026b, Prime Minister of Australia, 2026, National Heavy Vehicle Regulator, 2026, Department of Infrastructure Transport Regional Development Communications and the Arts, 2024); Market Shares (International Council on Clean Transportation, 2026a, International Council on Clean Transportation, 2026c, International Council on Clean Transportation, 2026b, International Council on Clean, 2026). NVES = New Vehicle Efficiency Standard; AFIR = Alternative Fuels Infrastructure Regulation.*

While the fuel price has gone up, the excise tax indexation was frozen from March 2001 to November 2014. In March 2026, the fuel excise tax is 52.6 cents per litre for HV diesel fuel on public roads. Nevertheless, the HV sector (trucks over 4.5 tonnes) is reimbursed for a segment of the fuel excise tax. The remaining portion of the fuel excise tax remains linked to road infrastructure financing via the Road User Charge (RUC). The RUC was 32.4 cents per litre in 2025–26. The RUC is applicable to every litre of fuel consumed by HVs (over 4.5 tonnes) that operate on public roadways, and constitutes part of the government's general revenue, which is used to finance a wide range of public services and infrastructure projects. The RUC rose in steps from about 19.6 cents per litre in 2006 to 21 cents in 2009 and about 25–26 cents over 2012–2021, and reached 27.2 cents in 2022–23, after which the Australian government introduced annual 6% increases from 2023–24 to 2025–26, ensuring that the fees imposed on HVs correspond to their negative externalities, including the impacts on road infrastructure (Australian Taxation Office, 2023). In late March 2026, however, in response to the Strait of Hormuz crisis, the Australian Government temporarily reduced fuel excise and set the HV RUC to zero from 1 April to 30 June 2026 (Australian Taxation Office, 2026a). The RUC was then set at 16.4 cents per litre from 1 July 2026 and restored to 32.4 cents per litre from 3 August 2026 (Australian Taxation Office, 2026b), and the next scheduled increase was deferred to 1 January 2027 (National Transport Commission, 2025). Environmental sustainability is a major part of the fuel excise tax discussion. By increasing the RUC, the freight sector is encouraged to reduce fuel usage, possibly transitioning to more fuel-efficient vehicles or ZEHVs.

Because Australian HVs remain in service for 12–16 years on average (Section 4), the fleet transition needs to start early to meet Australia's net-zero commitments. The majority of Australian freight operators are small to medium businesses, with more than 93 per cent of Australian trucking operators have a turnover of less than $2 million, and more than 98 per cent have 19 or fewer employees (Australian Bureau of Statistics, 2018). The high CAPEX of BETs/FCTs is a key adoption barrier because they do not have the necessary capital to purchase such vehicles. Therefore, various policy scenarios can be combined to reduce emissions and improve fuel efficiency, as well as reduce the TCO of BETs/FCTs. The remainder of this section compares Australia's HV emission and fuel-efficiency standards with those of other developed countries.

Fuel efficiency standards are regulations aimed at reducing vehicle fuel consumption and emissions, and they are typically set up through a combination of methods. One of the most common methods is establishing average targets for vehicle manufacturers based on their fleet-wide performance. This approach, often referred to as fleet averaging, requires manufacturers to meet a specified average CO2 emission or fuel consumption level across all vehicles they sell in a given market. Labelling is another method used to set up fuel efficiency standards, focusing on informing consumers to influence purchasing decisions and indirectly pressuring manufacturers to improve efficiency. Fuel efficiency labels display metrics such as litres per 100 km or $CO_2$ emissions, helping consumers compare vehicles. This is often accompanied by standards to create a synergistic effect, as noted by Mahlia et al. (2013).

In Australia, fuel efficiency standards are governed by the Australian Design Rules (ADR), a national standard established under the Road Vehicle Standards Act 2018, which sets mandatory requirements for vehicle performance, including emissions and fuel usage. Specific rules, such as ADR 79, ADR 80, and ADR 81, regulated aspects like exhaust emissions and fuel consumption to ensure compliance with environmental and efficiency goals. As noted above, the current HV rule, ADR 80/04, regulates noxious pollutants only, and Australia has no mandatory $CO_2$ or fuel-efficiency standards for HVs.

HV manufacturers in Australia offer a range of models, many of which are directly imported. According to the Commonwealth of Australia (2021), nearly half (47.7%) of all heavy-duty RTs and ATs supplied to the Australian market are produced locally to meet the specific needs of Australia's interstate road freight industry. Until ADR 80/04 took effect, these trucks were therefore not required to meet stricter global standards such as Euro VI. HVs built to comply with stricter international emissions standards (e.g. Euro VI) are equipped with advanced technologies, which increase production costs by approximately 3–5%. Advanced safety and emission features, which are mandatory in many overseas markets, are often offered as optional extras in Australia, adding to the cost (Commonwealth of Australia, 2021). Hence, the Australian consumers are less inclined to purchase models equipped with advanced safety and emissions systems due to their higher price points (Commonwealth of Australia, 2021).

Many developed countries, however, have mandatory $CO_2$ standards or targets for HVs (Table 1). For example, the EU has set targets for HVs starting from 2025, requiring the average $CO_2$ emissions of new HVs to fall by 15% in 2025–2029, 45% in 2030–2034, 65% in 2035–2039 and 90% from 2040, relative to 2019. Canada regulates the greenhouse gas emissions of new HVs under Phase 2 standards that tighten until model year 2027 and are aligned with the former US standards (Government of Canada, 2026), and Japan sets fuel-efficiency standards for HVs under its Top Runner programme, requiring an average improvement of about 13% by 2025 relative to 2015 (TransportPolicy.net). In the United States, by contrast, all federal greenhouse gas standards for HVs, including the Phase 3 standards adopted in 2024, were repealed in 2026 when the Environmental Protection Agency rescinded its 2009 Endangerment Finding, a decision that is being challenged in court (U.S. Environmental Protection Agency, 2026); state-level efforts, such as California's zero-emission truck rules, continue.

As a result, the TCO for ZEHVs in Australia is expected to be higher than in the European cases analysed in the TCO studies reviewed in Section 2 (e.g. (Noll et al., 2022)). While ICE HVs in Australia currently benefit from fuel tax credit arrangements that partially offset diesel excise costs, financial incentives for battery-electric and hydrogen fuel cell trucks remain relatively limited.

Focusing on Australia as a case study, this study examines how energy prices and push and pull instruments, namely RUC increases, carbon pricing and subsidies for the purchase and operating costs of BETs and FCTs, affect the uptake of ZEHVs, diesel use and fuel-tax revenue, and how the fiscal effects of these instruments bear on the funding of charging and refuelling infrastructure. Regulatory instruments such as $CO_2$ standards and ZEHV sales mandates, which distinguish the European packages in Table 1, are not modelled and are left for future work.

# METHODS

This section presents the SD modelling approach and formulations. A mixed-method approach has been adopted for the development of the SD model (Figure 1).

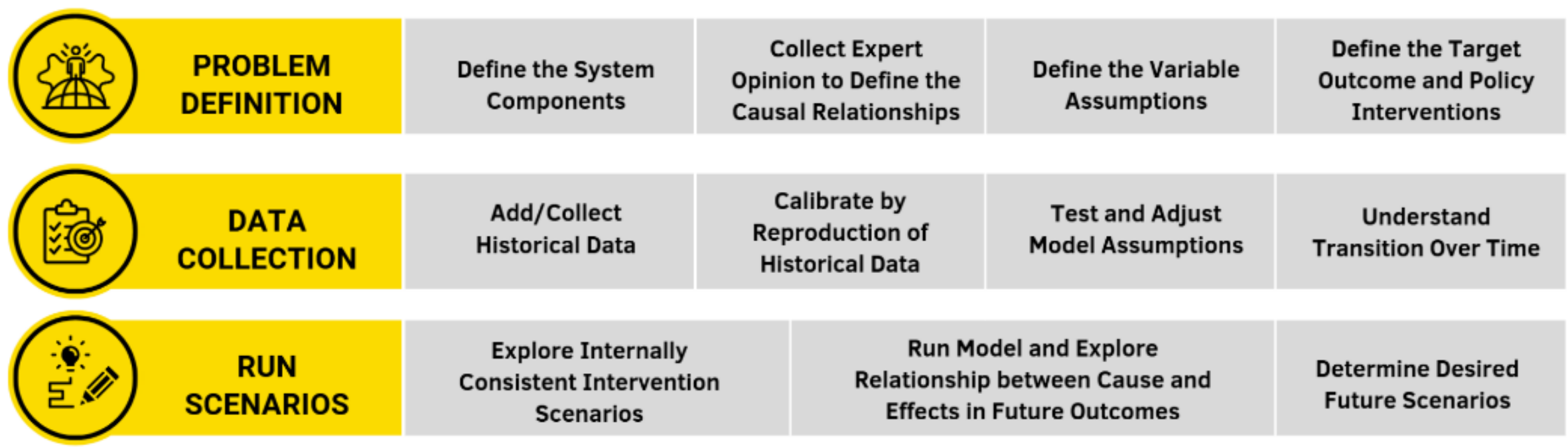

**Figure 1| Framework for the development and application of the SD model.** The figure illustrates the three-stage modelling process adopted in this study, comprising problem definition, data collection and calibration, and scenario analysis. The framework integrates system conceptualisation, expert-informed causal relationships, historical data calibration, model validation, and policy scenario testing to evaluate long-term freight transport transition pathways.

The model is calibrated against historical data for 2010–2023 (see Calibration below) and is simulated with Euler integration and a time step of one month. The input parameters driving the SD model are shown in Table 2. Most data are obtained from the Australian Bureau of Statistics (ABS) and the Bureau of Infrastructure and Transport Research Economics (BITRE), complemented by additional sector-specific sources as indicated.

**Table 2 | Summary of key datasets used as model inputs, including variables and data sources.**

| Description | Source(s) |
|---|---|
| Total number of vehicles | Motor Vehicle Census (Australian Bureau of Statistics, 2023) |
| GDP | Australian National Accounts: National Income, Expenditure and Product (Australian Bureau of Statistics, 2024) |
| Freight demand (tonne-km) | The Survey of Motor Vehicle Use (Australian Bureau of Statistics, 2020) |
| Australia transit road capacity | Infrastructure Statistics and Transport Yearbook (Bureau of Infrastructure, Transport and Regional Economics, 2022) |
| Average operation and maintenance costs (fuel costs excluded) and average revenue for various vehicle types | Truck OPEX & revenues (Freight Metrics, 2025) |
| Average fuel price | Estimated based on the retail petrol and diesel prices (Bureau of Infrastructure, Transport and Regional Economics, 2017) and the annual retail price for petrol and diesel (Australian Institute of Petroleum (AIP), 2022) and adjusted by model calibration. |
| Average electricity and hydrogen price and net change ratio | Estimated based on Speth et al. (2022)& Noll et al. (2022), and adjusted by model calibration. |
| Average fuel consumption per Tonne-km for various vehicle types | The Survey of Motor Vehicle Use (Australian Bureau of Statistics, 2020) |

Once the model reproduces past observed behaviour, we can evaluate policy interventions such as RUC increases, carbon pricing and subsidies for the purchase and operating costs of BETs and FCTs. The future simulation time horizon is from 2027 to 2050. Stella Architect software (isee systems, 2025) is used to build the model and the graphical user interface (GUI) dashboard. The subsystem diagram is presented in Figure 2.

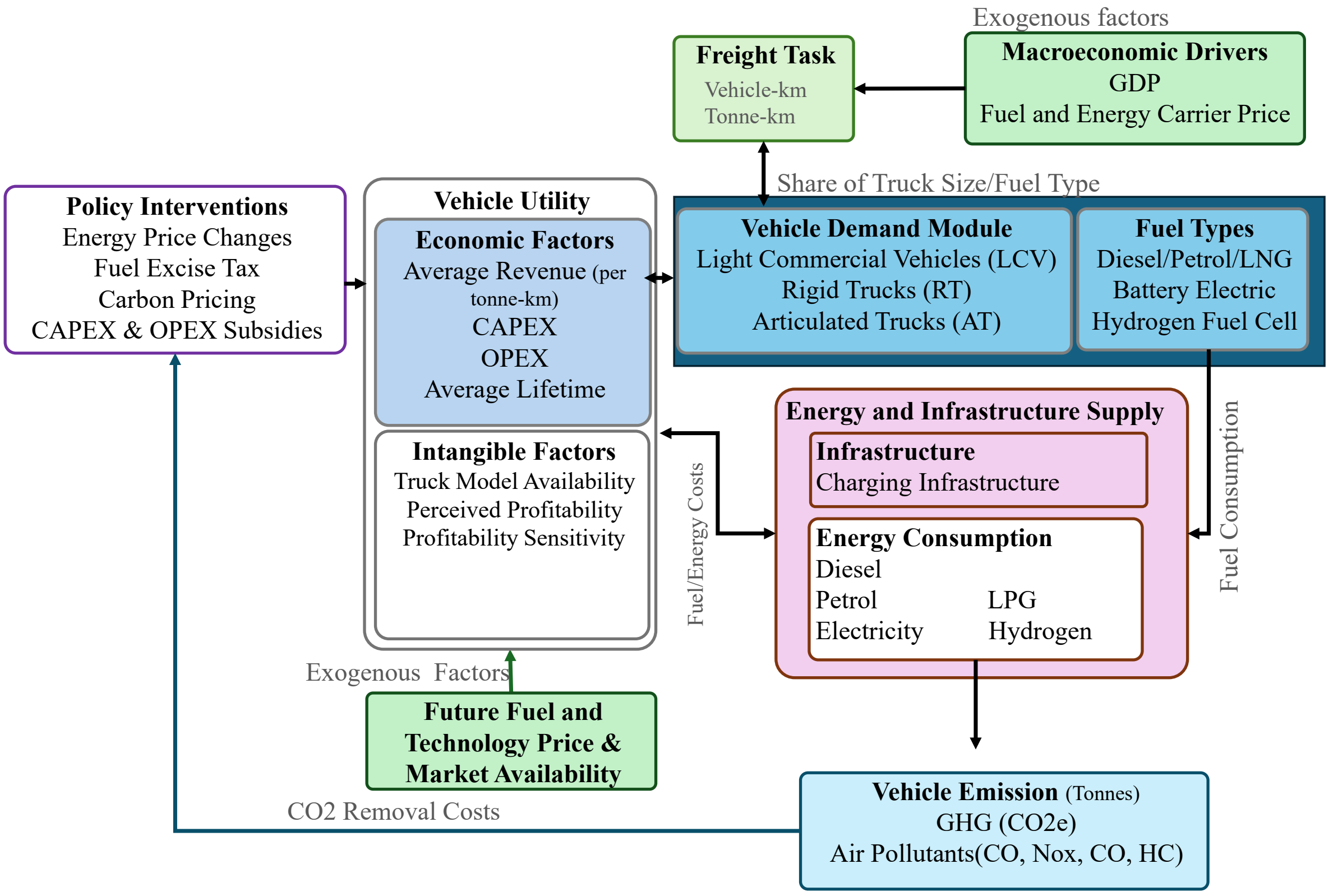

**Figure 2| Subsystem diagram.** At each monthly time step, the model simulates the evolution of the freight fleet as a function of macroeconomic drivers, including gross domestic product (GDP) and total freight demand measured in tonne-kilometres. Vehicle market share, representing the relative attractiveness of competing technologies, is determined through TCO calculations incorporating both CAPEX and OPEX. Historical freight activity and fuel consumption data are used to estimate macro-level tailpipe emissions from ICE trucks, including $CO_2$ and other air pollutants. The resulting externality costs are subsequently incorporated into the evaluation of taxation and decarbonisation policy scenarios. The expansion of the freight vehicle fleet is driven by new truck sales across multiple categories, all with different fuel types, collectively determining the system's total transport capacity. Growth in economic activity, as reflected by GDP, translates into higher freight demand (tonne-km), which in turn increases the volume of goods transported.

Figure 3 presents the CLD of the model, which comprises four reinforcing (R) and five balancing (B) loops. In the fleet sub-system, growth in freight tasks increases the number of new trucks purchased. As the fleet grows, its freight capacity rises and fewer new trucks are needed, which balances new sales (Balanced fleet growth, B). A larger fleet also retires more trucks each year, and these must be replaced (Fleet retirement, R). Expectations of future freight demand, shaped by current freight volumes, influence purchasing decisions and create a forward-looking effect on fleet expansion (Forward-looking fleet expansion, R). The freight carried generates ownership revenue that strengthens the attractiveness of each truck type (Ownership profitability, R), while the associated operating and maintenance costs raise the TCO and weaken it (Ownership cost, B). Together, these three reinforcing and two balancing loops sustain growth in freight capacity as long as demand continues to increase.

In the infrastructure sub-system, a policy will be tested where a share of fuel excise revenue is allocated to building charging and hydrogen refuelling infrastructure for BETs and FCTs. This investment improves the availability of charging and refuelling capacity and therefore the attractiveness of BETs and FCTs relative to diesel trucks. Higher adoption raises charging and refuelling demand and station utilisation, which drives further investment in the network (Battery/hydrogen refuelling network expansion, R). As more capacity is installed, average utilisation falls, which slows the construction of new capacity (Capacity expansion control, B). Conversely, if demand grows faster than capacity, the availability of the network falls and further adoption slows (Impact of battery/hydrogen infrastructure, B). Finally, as BETs and FCTs displace diesel trucks, diesel consumption and fuel excise revenue decline, eroding the very source of funding that supports the transition (Fuel-tax erosion, B).

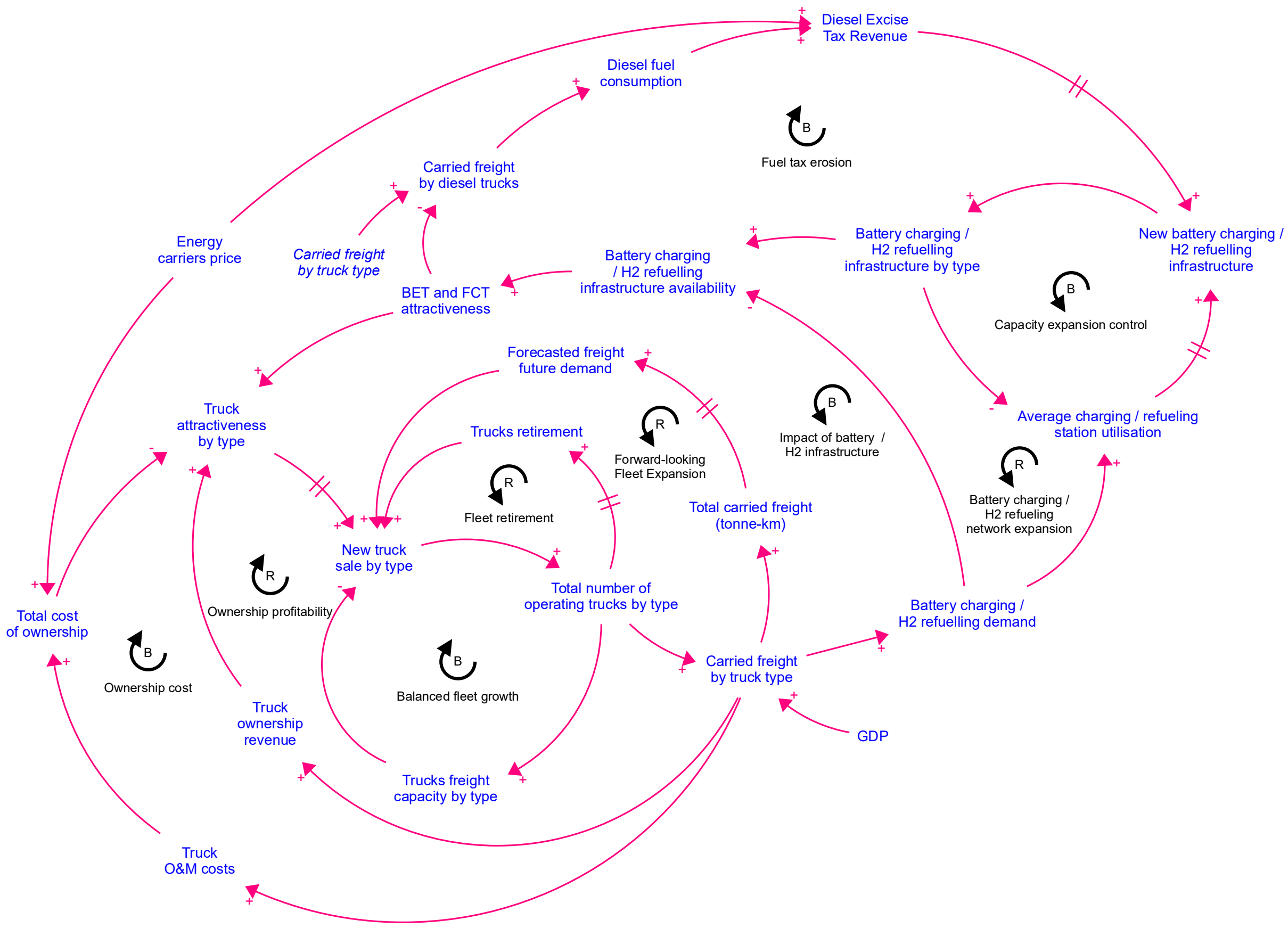

**Figure 3 | CLD of the model.** Arrows indicate causal relationships between variables, where a positive (+) sign denotes a change in the same direction and a negative (–) sign denotes a change in the opposite direction. Reinforcing loops (R) represent self-amplifying feedback mechanisms that drive system growth or decline, whereas balancing loops (B) represent stabilising feedback mechanisms that counteract change and promote system equilibrium. Delay symbols (||) on causal links indicate time lags between cause and effect, reflecting delayed system responses.

The following subsections present the model formulations, and the symbols used in this section are listed in Table 3.

**Table 3 | Notations**

| Symbol | Description |
|---|---|
| $TCO_{i,j}$ | Levelised total cost of ownership of vehicle type i with fuel j (AUD per tonne-km) |
| $CAPEX_{i,j}$ | Purchase cost of a new vehicle, incl. stamp duty and net of subsidies (AUD) |
| $SV_{i,j}$ | Salvage value at the end of the vehicle lifetime (AUD) |
| $OPEX_{i,j}$ | Annual operating cost (AUD per year) |
| $F_i$ | Annual freight task per vehicle (tonne-km per year) |
| $A_i$ | Annuity factor |
| $G_i$ | Growing-annuity factor for OPEX |
| $n_i$ | Vehicle lifetime (15 years for RTs, 12 years for ATs) |
| $r$ | Discount rate (6%) |
| $g$ | Expected growth of OPEX (2.5% per year) |
| $EC_{i,j}$ | Energy consumption per tonne-km (litres, kWh or kg) |
| $EP_j$ | Energy price |
| $MC_{i,j}$ | Maintenance cost per tonne-km |
| $I_{i,j}$ | Insurance cost per tonne-km |
| $CC_{i,j}$ | Carbon cost per tonne-km |
| $s_j$ | OPEX subsidy rate (energy and maintenance only) |
| $REG_i$ | Annual registration charge |
| $O_i$ | Other annual costs |
| $DW_i$ | Annual driver wage |
| $e_j$ | Net growth ratio of energy prices |
| $c_{i,j}$ | Net growth ratio of CAPEX |
| $\vartheta_{i,j}$ | Net growth ratio of insurance costs |
| $m_{i,j}$ | Net growth ratio of maintenance costs |
| $U_{i,j}$ | Attractiveness |
| $E_{i,j}$ | Commercial availability indicator |
| $M_{i,j}$ | Model availability ramp |
| $w_i$ | Segment-specific constant |
| $BCR_{i,j}$ | Benefit–cost ratio |
| $R_i$ | Revenue per tonne-km |
| $\gamma$ | Cost-sensitivity exponent (4.73) |
| $S_{i,j}$ | Perceived adequacy of the charging or refuelling network |
| $\tau_S$ | Adjustment time of perceived adequacy (5 years) |
| $RU_{i,j}$ | Relative attractiveness |
| $N_{i,j}$ | New vehicles purchased |
| $\hat{F}$ | Forecast total freight demand (tonne-km) |
| $K$ | Freight capacity of the fleet; Kret: capacity of retiring vehicles |
| $V_{i,j}$ | Number of vehicles in the fleet |
| $D_j$ | Annual charging or refuelling demand (kWh or kg) |
| $Q_{i,j}$ | Freight task carried by the fleet (tonne-km per year) |
| $Z_j$ | Installed capacity; Z*: capacity gap; z: completion rate of new capacity |
| $C_j$ | Effective capacity |
| $\varphi_j$ | Capacity factor of a station |
| $\tau_c$ | Construction time of a station (2 years) |
| $\tau_Z$ | Time to close the capacity gap (1 year) |
| $L$ | Station lifetime (25 years) |
| $B_j$ | Fuel-excise funding for station construction |
| $\kappa_j$ | Capital cost per unit of capacity |
| $\pi_j$ | Effect of station profitability on investment |
| $NPV_j$ | Net present value of a unit of capacity |
| $\sigma$ | Capital subsidy for stations (20%) |
| $Y_j$ | Annual revenue of a unit of capacity |
| $u_j$ | Utilisation of the network |
| $\Omega_j$ | Maximum annual throughput of a unit of capacity |
| $\omega_j$ | Station O&M cost as a share of revenue |
| $\delta_j$ | Energy distribution cost |

| Symbol | Description |
|---|---|
| $\lambda$ | Salvage value of a station (10% of capital cost) |
| $TE$ | Total energy consumption |
| $JR_j$ | Energy content of energy carrier j |
| $RUC$ | Road user charge per litre of diesel |
| $g_R$ | RUC baseline growth |
| $\rho$ | Annual increase rate of the RUC under the policy |
| $H$ | Heaviside step function |
| $t_s$ | Start year of the RUC increase |
| $v$ | Number of years of the RUC increase |

## TCO and Fleet Competition

The truck ownership model is developed based on the Polya technology competition model (Arthur et al., 1987). Truck relative attractiveness depends on: (i) the lifetime cost competitiveness of each truck type and fuel, measured by its levelised TCO relative to the revenue it earns; (ii) a segment-specific constant capturing non-cost attractiveness (e.g. suitability for the freight task, brand perception and operator familiarity); (iii) the perceived adequacy of charging and hydrogen refuelling infrastructure, which introduces path dependence through the infrastructure feedback loops; and (iv) the market availability of each technology.

Cost competitiveness is measured by a levelised TCO per tonne-km, i.e. the discounted lifetime cost of owning and operating a truck divided by its discounted lifetime freight task. TCO (per tonne-km) for vehicle type $i$ and fuel type $j$ is calculated as follows:

$$TCO_{i,j}(\mathrm{t}) = \frac{CAPEX_{i,j}(\mathrm{t}) - \frac{SV_{i,j}(\mathrm{t})}{(1+r)^{n_i}} + OPEX_{i,j}(\mathrm{t}) \cdot G_i}{F_i(\mathrm{t}) \cdot A_i} \quad (1)$$

where $CAPEX_{i,j}(\mathrm{t})$ is the purchase cost of a new vehicle in year $t$, including stamp duty (3% of the purchase price) and net of any purchase subsidy; $SV_{i,j}(\mathrm{t})$ is its salvage value at the end of its lifetime $n_i$, assumed to be 20% of the purchase price; $OPEX_{i,j}(\mathrm{t})$ is the annual operating cost; $F_i(\mathrm{t})$ is the annual freight task per vehicle (tonne-km), taken from the Survey of Motor Vehicle Use for each segment and held at its 2023 value thereafter; $r$ is the discount rate; $A_i = [1 - (1 + r)^{-n_i}]/r$ is the annuity factor that converts the annual freight task into a discounted lifetime freight task; and $G_i = [1 - ((1 + g)/(1 + r))^{n_i}]/(r - g)$ is the corresponding growing-annuity factor for OPEX that operators expect to rise by $g$ = 2.5% per year. Because both the numerator and the denominator are discounted, Eq. (1) gives the constant price per tonne-km at which an operator would recover all lifetime costs.

The annual OPEX is calculated as follows:

$$OPEX_{i,j}(\mathrm{t}) = F_i(\mathrm{t})\left[\left(1 - s_j(\mathrm{t})\right)\left(EC_{i,j}EP_j(\mathrm{t}) + MC_{i,j}(\mathrm{t})\right) + I_{i,j} + CC_{i,j}(\mathrm{t})\right] + REG_i + O_i + DW_i \quad (2)$$

where $EC_{i,j}$ is the energy consumption per tonne-km (litres, kWh or kg), $EP_j(\mathrm{t})$ the energy price, $MC_{i,j}(\mathrm{t})$ the maintenance cost per tonne-km, $I_{i,j}$ the insurance cost per tonne-km, $CC_{i,j}(\mathrm{t})$ the carbon cost per tonne-km (zero unless carbon pricing applies), $s_j(\mathrm{t})$ the OPEX subsidy rate for ZEHVs, which applies to energy and maintenance costs only, $REG_i$ the annual registration charge, $O_i$ other annual costs (e.g. tyres) and $DW_i$ the annual driver wage.

In the literature, a discount rate of 5-8% was considered for private and commercial vehicle ownership calculations (Wang et al., 2024). It reflects moderate assumptions about inflation, cost of capital, and the risk-free rate for long-term investments. Some studies on electric vehicle ownership have used discount rates in the 5-6% range when assessing the lifecycle costs, reflecting relatively low financial risk and the stable long-term growth of EV adoption (de Albuquerque Felizola Romeral and Zancul, 2025). Studies evaluating the TCO for HVs may use slightly higher rates (around 6-8%) due to the higher risks and costs associated with these assets (Norris and Escher, 2017, Meszler, 2019). In this study, we consider the discount rate of 6%. Because small operators, which dominate the Australian trucking industry, are likely to face a higher cost of capital, we also tested discount rates of 10% and 15%. At 15%, the share of CAPEX in the levelised TCO of a battery AT in 2030 rises from 15% to 22% (from 8% to 14% for a battery RT) and the TCO parity of battery trucks with diesel trucks is delayed by four to five years, yet the cumulative ZEHV share of new RT and AT sales in 2050 falls by only 1.5 percentage points (0.6 percentage points at 10%). OPEX therefore still accounts for more than three-quarters of the TCO of BETs, so the greater leverage of OPEX instruments (H3) does not depend on the choice of discount rate, although the gap between OPEX and CAPEX instruments narrows as the discount rate rises.

The Australian Motor Vehicle Census data suggest that the average age of Australian HVs is as follows: light rigid trucks (11 years), heavy rigid trucks (16 years); articulated trucks (12 years); and non-freight trucks (14 years)(Australian Bureau of Statistics, 2023). The Australian Tax Office considers 12 and 15 years for LCVs and rigid trucks, respectively (Australian Taxation Office). In this study, we assume 15 and 12 years for vehicle lifetime for RT and ATs, respectively.

Vehicle depreciation and resale value depend on various factors such as mileage, powertrain, brand, vehicle features, and conditions. Rout et al. (2022) assumed RV of 50% for fossil-fuelled trucks after the first two years of ownership, with the rates slowing towards the maximum vehicle lifetime of 16 years, and after 12 years, the RV drops to 20%, where it remains until the vehicle lifetime. The same rate was applied to BETs and FCTs by Rout et al. (2022). Another study by Carpenter and Antich Arnau (2023) suggests minimal difference in depreciation across powertrains. In this study, the salvage value is assumed to be 20% of the purchase price at the end of the vehicle lifetime (adopted from (Rout et al., 2022, Australian Taxation Office, 2025)). It should be noted that a sensitivity analysis showed that halving the BET/FCT salvage value from 20% to 10% of the purchase price raises their levelised TCO by about 0.8% (ATs) and 0.4% (RTs) in 2030 and, with $\gamma = 4.73$ (Eq. 4), lowers their attractiveness relative to diesel trucks by about 3.6% and 1.7%, respectively.

In the literature, different values of average energy consumption rates have been presented for various truck sizes (e.g. (Noll et al., 2022, Rout et al., 2022, Burke and Miller, 2020, Speth et al., 2022)). In this paper, we considered the cost values and growth rates presented in Appendix B.

Growth ratios were also considered as follows:

$$\frac{\mathrm{d}\boldsymbol{EP_j}}{\mathrm{dt}} = \boldsymbol{e_j} \cdot \boldsymbol{EP_j}(\mathbf{t}) \; ; \; \frac{\mathrm{d}\boldsymbol{CAPEX_{i,j}}}{\mathrm{dt}} = \boldsymbol{c_{i,j}} \cdot \boldsymbol{CAPEX_{i,j}}(\mathbf{t}) \; ; \; \frac{\mathrm{d}\boldsymbol{I_{i,j}}}{\mathrm{dt}} = \boldsymbol{\vartheta_{i,j}} \cdot \boldsymbol{I_{i,j}}(\mathbf{t}) \; ; \; \frac{\mathrm{d}\boldsymbol{MC_{i,j}}}{\mathrm{dt}} = \boldsymbol{m_{i,j}} \cdot \boldsymbol{MC_{i,j}}(\mathbf{t}) \tag{3}$$

where $e_j$ is the average energy price net growth ratio, $c_{i,j}$ is the average CAPEX net growth ratio; $\vartheta_{i,j}$ is the average insurance cost net growth ratio; $m_{i,j}$ is the average maintenance net growth ratio. All these net growth rates are considered positive for diesel and petrol, and negative for electricity and hydrogen, as shown in Appendix Table B1.

The attractiveness of truck type *i* with fuel *j* combines its cost competitiveness with non-cost factors:

$$U_{i,j}(\mathrm{t}) = E_{i,j}(\mathrm{t}) \cdot M_{i,j}(\mathrm{t}) \cdot w_i \cdot \left(\frac{BCR_{i,j}(\mathrm{t})}{BCR_{i,D}(2010)}\right)^{\gamma} \cdot S_{i,j}(\mathrm{t}) \tag{4}$$

where $BCR_{i,j}$(t) = $R_i$/$TCO_{i,j}$(t) is the benefit–cost ratio, i.e. the revenue per tonne-km $R_i$ (Freight Metrics, 2025) divided by the levelised TCO (Eq. 1). Normalising by the 2010 benefit–cost ratio of the diesel truck of the same segment, $BCR_{i,D}$(2010), makes this term equal to one for every diesel segment in 2010 and, because $R_i$ is common to all fuels of a segment, reduces it to the cost ratio $TCO_{i,D}$(2010)/$TCO_{i,j}$(t). The exponent $\gamma$ measures how strongly attractiveness responds to cost differences. For example, with the calibrated value of 4.73, a 10% lower levelised TCO raises attractiveness by about 65%. $w_i$ is a segment-specific constant capturing non-cost attractiveness; $E_{i,j}$(t) is an indicator equal to one once technology *j* is commercially available in segment *i* (2027 for battery RTs and ATs, and 2030 for fuel cell trucks); $M_{i,j}$(t) is a logistic model-availability ramp for battery and fuel cell vehicles and equals one otherwise; and $S_{i,j}$(t) is the perceived adequacy of the charging or refuelling network (one for fossil-fuel trucks), which adjusts gradually to the ratio of capacity to demand:

$$\frac{\mathrm{d}S_{i,j}(\mathrm{t})}{\mathrm{d}t} = \frac{\min\left[1, \frac{C_j(\mathrm{t})}{D_j(\mathrm{t})}\right] - S_{i,j}(\mathrm{t})}{\tau_S} \tag{5}$$

where $C_j$(t) is the charging ($j$ = electricity) or refuelling ($j$ = hydrogen) capacity, $D_j$(t) is the corresponding demand from battery or fuel cell trucks, and $\tau_S$ = 5 years is the time operators take to perceive and trust the adequacy of the network; because it cannot be estimated from historical data, it is varied between 2 and 10 years in the sensitivity analysis (Section 5). An initial test of $\tau_S$ =20 years implied a perceived adequacy of only 0.68 in 2050 even with ample capacity. If charging and refuelling capacity exceeds demand, so that min[1, $C_j(t)/D_j(t)$] = 1 and Eq. (5) reduces to $S_{i,j}(t) = 1 - \exp[-(t - t_0)/\tau_S]$, where $t_0$ is the year in which technology *j* enters the market. Perceived adequacy then follows an exogenous confidence ramp (0.45 in 2030, 0.80 in 2035 and 0.99 in 2050 for battery trucks), and $\tau_S$, rather than the physical network, governs the lag between cost parity and adoption unless capacity falls short of demand. Relative attractiveness, which allocates new vehicle purchases among truck types and fuels, is then:

$$RU_{i,j}(t) = \frac{U_{i,j}(t)}{\sum_{(i,j)} U_{i,j}(t)} \tag{6}$$

New trucks are purchased to replace retiring vehicles and to close the gap between forecast freight demand and the capacity of the existing fleet, and are allocated according to relative attractiveness:

$$N_{i,j}(\mathrm{t}) = \max\left[0, \hat{F}(\mathrm{t}) - K(\mathrm{t}) + K^{\mathrm{ret}}(\mathrm{t})\right] \cdot \frac{RU_{i,j}(\mathrm{t})}{F_i(\mathrm{t})} \tag{7}$$

where $\hat{F}(t)$ is the one-year-ahead forecast of total freight demand (a trend extrapolation with a three-year averaging time), $K(t) = \Sigma V_{i,j}(\mathrm{t})F_i(\mathrm{t})$ is the freight capacity of the current fleet of $V_{i,j}$ vehicles, and $K^{ret}(t)$ is the capacity of the vehicles retiring or scrapped in year *t*. Because all segments compete for the same freight task, relative attractiveness determines both the fuel mix within a segment and the division of new capacity between segments.

## Charging Infrastructure

Charging and hydrogen refuelling capacity is built in response to forecast demand and station profitability, and the perceived adequacy of the resulting network enters truck attractiveness through Eq. (5). The annual energy demand of BETs ($j$ = electricity) and FCTs ($j$ = hydrogen) is determined by the freight task of the fleet in service:

$$D_j(\mathrm{t}) = \sum_{\mathrm{i}\in\{\mathrm{RT,AT}\}} EC_{i,j} \cdot Q_{i,j}(\mathrm{t}) \tag{8}$$

where $Q_{i,j}(t)$ is the freight task (tonne-km per year) carried by all vehicles of type $i$ and fuel $j$ in service and $EC_{i,j}$ is their energy consumption per tonne-km. Station operators plan new capacity to close the gap between the capacity needed for the demand expected once construction is complete and the installed capacity of existing stations:

$$Z_j^*(\mathrm{t}) = \max\left[0, \frac{\hat{D}_j(t+\tau_c)}{\varphi_j^2} - Z_j(\mathrm{t})\right] \quad (9)$$

where $Z_j^*(t)$ is the capacity gap, $\hat{D}_j(t + \tau_c)$ is a trend forecast of demand $\tau_c$ = 2 years ahead (the construction time of a station), $\varphi_j$ is the capacity factor of a station (0.885 for charging and 0.92 for hydrogen refuelling; Table B3) and $Z_j(t)$ is the installed capacity. Operators therefore aim for an installed capacity of $\hat{D}_j/\varphi_j^2$, which adds a planning margin of $1/\varphi_j$ (13% for charging and 9% for hydrogen refuelling) to the capacity needed to serve the forecast demand at the station capacity factor, $\hat{D}_j/\varphi_j$. The effective capacity used in Eq. (5) is $C_j(t) = \varphi_j Z_j(t)$.

Installed capacity accumulates completed capacity, $z_j(t)$, and retires at the end of the station lifetime $L$ = 25 years:

$$\frac{\mathrm{d}Z_j(\mathrm{t})}{\mathrm{dt}} = z_j(\mathrm{t}) - z_j(t-L) \quad (10)$$

New capacity is completed after a first-order construction delay:

$$\frac{\mathrm{d}z_j(\mathrm{t})}{\mathrm{dt}} = \frac{\frac{B_j(\mathrm{t})}{\kappa_j} + \frac{Z_j^*(\mathrm{t})\cdot\pi_j(\mathrm{t})}{\tau_Z} - z_j(\mathrm{t})}{\tau_c} \quad (11)$$

where $B_j(t)$ is the funding allocated to station construction from fuel-excise revenue when this policy option is enabled (Section 5), $\kappa_j$ is the capital cost per unit of capacity, $\tau_Z$ = 1 year is the time over which operators plan to close the capacity gap and $\pi_j(t)$ is the effect of station profitability on investment:

$$\pi_j(\mathrm{t}) = \max\left[1, 2 - \exp\left(-\frac{NPV_j(\mathrm{t})}{\kappa_j}\right)\right] \quad (12)$$

$\pi_j(t)$ equals one when stations are unprofitable, so that capacity simply follows demand, and approaches two as the net present value of a unit of capacity, $NPV_j(t)$, rises relative to its capital cost:

$$NPV_j(\mathrm{t}) = -(1-\sigma)\kappa_j + \sum_{k=\tau_c+1}^{\tau_c+L} \frac{Y_j(\mathrm{t})(1-\omega_j) - \delta_j(\mathrm{t})}{(1+r)^k} + \frac{\lambda\kappa_j}{(1+r)^{\tau_c+L}} \quad (13)$$

where $\sigma$ = 20% is the capital subsidy, $Y_j(t) = EP_j(t)u_j(t)\Omega_j$ is the annual revenue of a unit of capacity, $u_j(t) = D_j(t)/C_j(t)$ is the utilisation of the network, $\Omega_j$ is the maximum annual throughput of a unit of capacity, $\omega_j$ is the operating and maintenance cost of a station as a share of its revenue (22.5% for charging and 37.5% for hydrogen refuelling), $\delta_j(t) = 0.15EP_j(t)\Omega_j$ is the annual energy distribution cost of a unit of capacity, i.e. 15% of the energy price charged on its maximum annual throughput, $\lambda$ = 10% is the salvage value of a station and $r$ = 6% is the discount rate (Table B3).

## Tailpipe Gas Emission and Energy Consumption Estimation

This study utilises the Vehicle Census Data, published by the Australian Bureau of Statistics (2023), which provides the historical vehicle age and vintage year for different vehicle categories. The tailpipe gas emissions were then estimated based on the vehicle type and vintage year (which translates into various Euro standards), using the emission factors for different vehicle categories and different Euro standards. Since the Euro standards do not include $CO_2$ emissions, we separately estimate the tailpipe $CO_2$ emissions based on fuel consumption. It is widely accepted to assume that one litre of petrol will produce about 2.3 kg of $CO_2$, and 1 litre of diesel will produce about 2.7 kg of $CO_2$. Using the historical data on the total fuel consumption, we calculate total annual $CO_2$ emissions for different vehicle classes over 2010–2023. Total energy consumption, $TE(t)$, is calculated as follows:

$$TE(\mathrm{t}) = \sum_j JR_j \cdot \sum_i EC_{i,j}(\mathrm{t}) \cdot Q_{i,j}(\mathrm{t}) \tag{14}$$

where $Q_{i,j}(t)$ is the freight task (tonne-km per year) carried by vehicles of type $i$ and fuel $j$, $JR_j$ is the energy content of energy carrier $j$ (MJ per litre, kWh or kg) and $EC_{i,j}(t)$ is the energy consumption per tonne-km.

## Fuel Excise Tax

In this study we test two scenarios (Sncearios E and H in Section 5) where a percentage of the diesel excise tax goes toward investments in charging stations for ZEHVs, as well as road repair and road-related infrastructure costs. Therefore, raising the fuel excise tax rate to reflect the real environmental cost of fuel consumption might incentivise transitioning to ZEHVs.

With the increase in fossil fuel excise tax, the fossil fuel OPEX are increased, the profitability of ICE trucks is reduced, and, in Scenarios E and H, part of the tax income funds investment in charging infrastructure. The policy acts on the RUC, the part of the diesel excise that HVs pay after fuel tax credits, which is modelled as a stock that grows at a compound rate:

$$\frac{dRUC(t)}{dt} = [g_R(t) + (\rho - g_R(t))H(t - t_s)H(t_s + v - t)]RUC(t) \tag{15}$$

where RUC($t$) is the road user charge per litre of diesel, $g_R(t)$ is its baseline growth rate (1.5% a year before 2023, 6% over 2023–2026 and 2.5% thereafter), $\rho$ is the annual increase rate under the policy, $H(\cdot)$ is the Heaviside step function, $t_s$ is the start year of the increase and $v$ is the number of years over which it is applied. Because the increase compounds at the monthly time step of the model, it multiplies the RUC by $(1 + \rho/12)^{12v}$, i.e. by 2.1 for $\rho$ = 7.5% and by 19.4 for $\rho$ = 30% over $v$ = 10 years, taking it from 0.33 AUD per litre in 2027 to 0.69 and 6.33 AUD per litre in 2037, respectively.

## Calibration

The model was calibrated against annual data for 2010–2023 in two steps (Oliva, 2003, Sterman, 2000). First, parameters that can be observed directly were set from data rather than estimated. The initial fleet by segment and fuel (Motor Vehicle Census (Australian Bureau of Statistics, 2023)), the freight task per vehicle and the energy use per tonne-km by segment (Survey of Motor Vehicle Use (Australian Bureau of Statistics, 2020)), vehicle prices, OPEX and their growth rates (Appendix B), and the trend in diesel prices, for which the long-run growth rate in Appendix Table B1 was scaled by a factor of 0.742 so that the simulated price reproduces the observed rise from 1.30 AUD per litre in 2010 to about 2.0 AUD per litre in 2023. Second, the behavioural parameters of the attractiveness function (Eq. 4), which cannot be observed directly, were estimated by minimising the discrepancy between simulated and observed behaviour.

The objective function was the weighted sum of the time-integrated mean absolute percentage errors (MAPE) of 13 historical series over 2010–2023 for the freight task, the share of the freight task and the number of vehicles in each of the three segments, the diesel consumption of RTs and ATs, and the diesel and petrol consumption of LCVs. Because the analysis in this paper focuses on HVs, the RT and AT series were given a weight of one (0.5 for fleet size) and the LCV series a weight of 0.25. The search used the Powell (BOBYQA) optimiser in Stella Architect (isee systems, 2025) with a convergence tolerance of $10^{-8}$ and five additional starting points spread across the parameter space to guard against local optima. Calibration was also used as a test of model structure (Oliva, 2003).

Because ZEHV sales in Australia were negligible during the calibration period, $\gamma$ is identified from the response of the diesel fleet to changes in relative costs between segments, notably the larger effect of rising diesel prices on ATs, for which fuel accounts for about a third of OPEX, than on RTs, for which it accounts for about a tenth. Its application to the choice between zero-emission and diesel trucks is therefore an extrapolation, which we discuss as a limitation in Section 6.

## Validation

The model was evaluated using a combination of structural and behaviour-reproduction tests, following standard practice in SD modelling (Barlas, 1996, Sterman, 2000). Structural assessment comprised verification of the model structure and parameters against the literature and the data sources in Table 2 and Appendix B, and the use of calibration as a test of model structure (Oliva, 2003). Behaviour reproduction was assessed by comparing simulated and observed annual diesel consumption (million litres), freight task (billion tonne-km) and number of trucks (thousands) for RTs and ATs over 2010–2023 (Figure 4). The model reproduces all six series closely. MAPEs range from 0.9% to 2.2% (maximum annual error 5.6%), and Theil's inequality coefficients range from 0.006 to 0.013, well within the error levels commonly regarded as acceptable for strategic policy models.

For ATs, which carry about 80% of the RT and AT freight task, the MAPE is 2.0% for diesel consumption, 2.2% for the freight task and 1.8% for the number of trucks ($R^2$ of 0.73, 0.93 and 0.95, respectively). For RTs, the corresponding MAPEs are 2.1%, 2.0% and 0.9% ($R^2$ of 0.96, 0.95 and 0.99). The Theil decomposition attributes most of the remaining error to unsystematic, point-to-point variation (the share of the error due to imperfect point-by-point correlation, $U^C$, between 0.41 and 0.85), with small contributions from systematic bias (the share due to a difference in means, $U^M \leq 0.31$; mean bias between 0.0% and +1.3%) and from unequal variation (the share due to a difference in standard deviations, $U^S \leq 0.37$), following (Sterman, 1984). The simulated trajectories are smoother than the observed ones, which, for example, show a temporary slowdown in AT freight activity over 2015–2018 that a trend-driven model is not designed to reproduce. The model also reproduces the division of the freight task between segments. The simulated RT and AT shares deviate from the observed shares by 0.45 and 0.48 percentage points on average (maximum 1.3 percentage points). The LCV series, which received a lower weight in calibration, are reproduced with MAPEs of 1.7–2.9%, although the model does not capture the 12% increase in the LCV fleet between 2021 and 2023, which was driven largely by non-freight use. The diesel price is represented as a calibrated trend that reproduces the 2010–2023 increase but not the year-to-year volatility of prices (MAPE 15.7%).

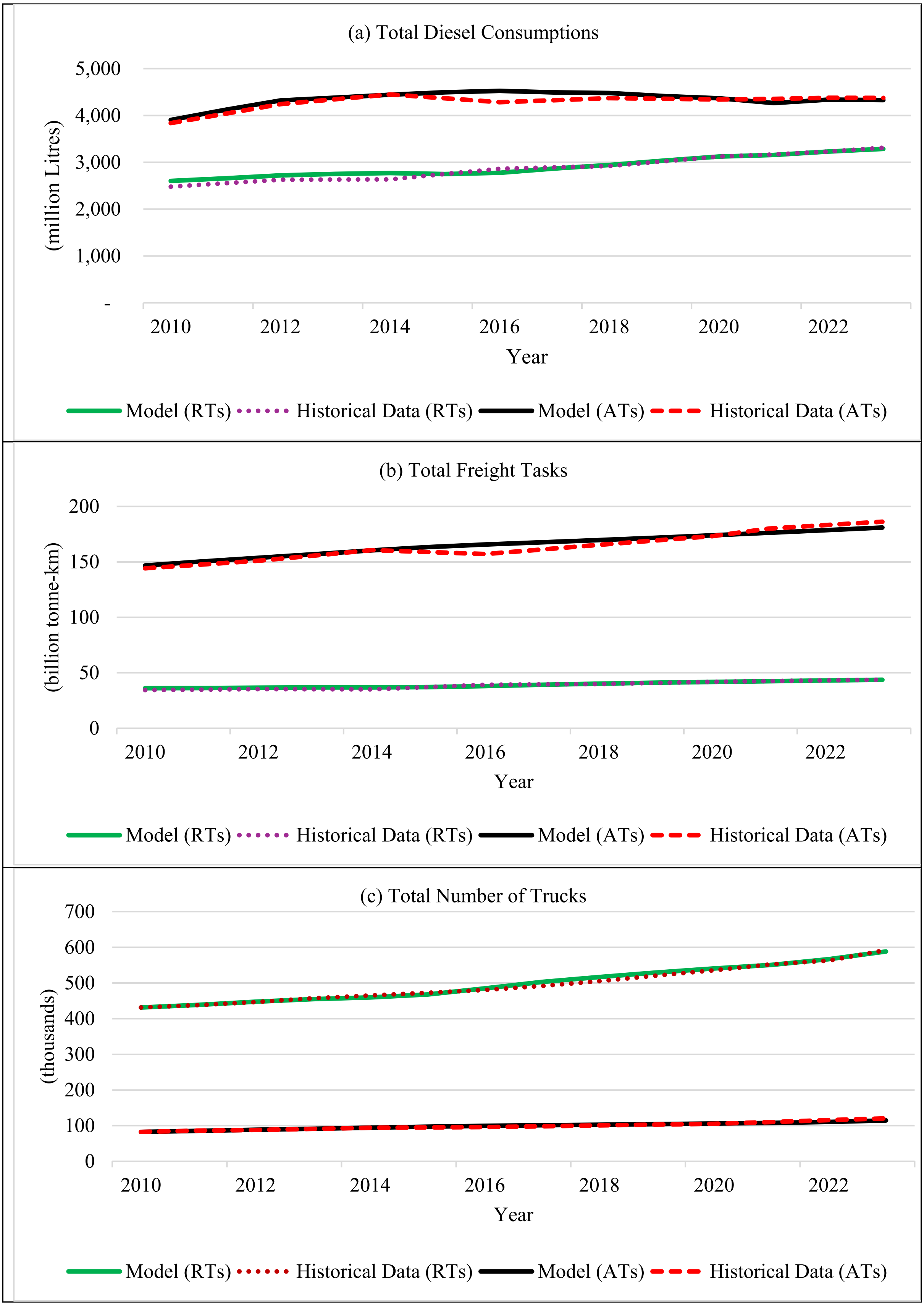

**Figure 4 | Model Validation Results.** The continuous line represents the model and the dashed line represents the historical data.

## Scenario Design

We simulate two sets of scenarios over 2027–2050 and compare each with the baseline, which continues current trends in energy prices, vehicle prices and the RUC. The energy-price scenarios (A–D; Table 4) test how the transition responds to the size, timing and persistence of changes in relative energy prices. They are motivated by the recent volatility of global fuel markets. Heightened geopolitical tensions in the Middle East have underscored the susceptibility of oil prices to sudden supply disruptions, and fuel shortages in Australia add uncertainty to relative fuel costs. The policy scenarios (E–I; Table 5) test push and pull instruments under baseline energy prices. Adoption is measured mainly by the cumulative share of BETs and FCTs in new RT and AT sales since 2027 which, unlike the share of annual sales, reflects how many ZEHVs enter the fleet over the projection period. The shares of the fleet and of the freight task, diesel use, emissions and fuel-tax revenue are reported alongside it, and each scenario is tested under alternative behavioural assumptions to test the robustness of the results (Section 5).

**Table 4: Energy-price scenarios (A–D)**

| Scenarios | Energy Carrier Price |
|---|---|
| Reference Scenario (Baseline Trends) | Calibrated trends: petrol and diesel about +3.1%/yr (about +3.3%/yr including RUC); electricity −0.9%/yr and hydrogen −0.6%/yr from 2030 (phased in from 2026); RUC +2.5%/yr from 2026 |
| Scenario A: Moderate Fossil Increase with Clean Energy Shift | Fossil fuels +10%/yr and Electricity & Hydrogen −10%/yr (from 2027 onward), other parameters same as baseline |
| Scenario B: Extreme Fossil Price Shock (stress test) | Fossil fuels +50%/yr in 2027–2032, then baseline growth; other parameters same as baseline |
| Scenario C: Strong Clean Energy Shift (stress test) | Electricity & Hydrogen −50%/yr (from 2027 onward); other parameters same as baseline |
| Scenario D: Baseline Trends with a 6%/yr increase of RUC | Same energy price trend as baseline, RUC increase +6%/yr (from 2023 onwards) |

Scenario A represents a persistent, two-sided shift in relative energy prices where fossil fuel prices rise, and electricity and hydrogen prices fall, at a continuous rate of 10% per year from 2027, so that by 2050 diesel costs about 4.6 times its baseline price (21.9 versus 4.8 AUD per litre) and electricity and hydrogen are about 90% cheaper (reflecting an assumption about overnight charging using the saved solar energy at the warehouses). Scenarios B and C are stress tests rather than price forecasts. Scenario B imposes an extreme fossil-fuel price shock where diesel prices rise at a continuous rate of 50% per year over 2027–2032, reaching about ten times the baseline price, and then resume the baseline trend, so that diesel remains about ten times more expensive than in the baseline in 2050. Scenario B therefore front-loads its price increase, and its diesel price ends at about twice that of Scenario A. In Scenario C, electricity and hydrogen prices fall steeply, by at a continuous rate of 50% per year (about 40% a year) from 2027, so that both become almost free by the mid-2030s, representing, for example, depot charging from on-site solar generation and storage. It should be highlighted that Scenario C is an upper-bound stress test, given network charges alone would prevent grid electricity from becoming almost free. Scenario D retains the baseline energy prices but raises the RUC by 6% per year from 2023 (compared with 2.5% per year from 2026 in the baseline), a realistic policy benchmark that more than doubles the RUC and lifts the diesel pump price by 16% relative to the baseline by 2050. These energy price scenarios are designed to represent the effects of the size, timing and persistence of energy-price signals on the HV fleet transition.

The policy scenarios, however, hold energy prices at their baseline trends, and instead isolate the effects of RUC increases, carbon pricing, CAPEX and OPEX subsidies, individually and in combination. They are designed to disentangle push and pull measures (Table 5).

**Table 5: Policy scenarios (E–I).**

| Policy Scenarios | RUC | Carbon Pricing Tax | CAPEX Subsidy and Price Reduction | OPEX Subsidy |
|---|---|---|---|---|
| Baseline | 6%/yr increase in 2023-26, then fixed at 2.5%/yr | None | No subsidy; purchase prices of BETs and FCTs fall by 1.34%/yr and 1.35%/yr, diesel trucks rise by 0.54%/yr | None |
| Scenario E: Push & Pull | +7.5%/yr RUC increase over 2027–2037 | 70 AUD per tonne of $CO_2$ from 2027, rising by 7 AUD per tonne each year (diesel RTs and ATs) | A purchase subsidy from 2027 covering 60% of the price premium over an equivalent diesel truck, as in Norway's Enova programme (Table 1); it phases itself out as the premium shrinks. | 20% reduction in the energy and maintenance costs of BETs and FCTs from 2027 onwards |
| Scenario F: Pull Only (CAPEX) | No change from Baseline Scenario | | A purchase subsidy from 2027 covering 60% of the price premium over an equivalent diesel truck, as in Norway's Enova programme (Table 1); it phases itself out as the premium shrinks. | None |
| Scenario G: Pull Only (OPEX) | No change from Baseline Scenario | | None | 20% reduction in the energy and maintenance costs of BETs and FCTs from 2027 onwards |
| Scenario H: Push Only (RUC) | 7.5%/yr RUC increase over 2027–2037 | None | None | None |
| Scenario I: Push Only (Carbon Pricing) | No change from Baseline Scenario | 70 AUD per tonne of $CO_2$ from 2027, rising by 7 AUD per tonne each year (diesel RTs and ATs) | None | None |

Scenario H is a push-only measure that raises the RUC by 7.5% per year over 2027–2037, from 0.33 to 0.69 AUD per litre, so that HVs pay the full diesel excise from 2037 onwards, which is equivalent to removing the fuel tax credit for on-road HVs. It lifts the diesel pump price by about 9% relative to the baseline in 2037 and 8% in 2050. To show how adoption responds to the level of the RUC, Scenario H was also run with annual increases of 15%, 20% and 30%. In Scenario H and its variants, as in Scenario E, 20% of fuel-tax revenue is allocated to the construction of charging and refuelling capacity; this allocation is switched off in the baseline and all other scenarios. Scenario I is a push-only carbon price of 70 AUD per tonne of $CO_2$ from 2027, rising by 7 AUD per tonne each year and applied to the diesel used by RTs and ATs, equivalent to about 0.19 AUD per litre of diesel in 2027 and 0.62 AUD per litre in 2050. Scenario G is a pull-only OPEX measure that reduces the energy and maintenance costs of BETs and FCTs by 20% from 2027, representing, for example, preferential electricity tariffs or lower maintenance costs of BETs/FCTs. Scenario F is a pull-only CAPEX measure. From 2027, a purchase subsidy covers 60% of the price premium of a BET or FCT over an equivalent diesel truck, the maximum share offered by Norway's Enova programme (Table 1). Because BET and FCT prices fall while diesel truck prices rise (Appendix Table B1), the subsidy phases itself out as the premium shrinks. It is worth about 51,700 AUD per battery AT (13% of its price) and 26,900 AUD per battery RT (19%) in 2027 and ends in about 2040 and 2047, respectively, whereas fuel cell ATs and RTs receive about 110,000 AUD (23%) and 56,100 AUD (29%) in 2027 and remain subsidised, at a declining rate, until 2050. Scenario E combines all of these instruments at the same levels as in Scenarios F–I. Comparing Scenarios F–I isolates the individual instruments, while comparing Scenario E with the sum of their effects indicates how the instruments interact.

# RESULTS

This section presents the summary of results, including the elasticity calculations, cost parity, size, timing and persistence in the adoption of ZEHVs, push versus pull scenarios, rigid versus articulated trucks, infrastructure dependencies, the outcomes of policies in terms of total diesel consumption, CO2 emissions, and fuel tax revenue, and sensitivity analysis of the model parameters.

## A Lower Lifetime Cost without Rapid Adoption in the Baseline Scenario

In the baseline, ZEHVs account for 20% of cumulative new RT and AT sales by 2030, 39% by 2035, 49% by 2040 and 60% by 2050 (Figure 5), and their share of annual sales rises from 31% in 2030 to 59% in 2035, 68% in 2040 and 77% in 2050. FCTs, which become commercially available from 2030, account for 17 percentage points of the 2050 cumulative share. These shares rise without policy support when they become readily available in the market because, ZEHV lower OPEX outweighs their higher purchase price, so their levelised TCO reach to the cost parity of diesel trucks. In 2030, the levelised TCO of a battery AT is 11% below that of a diesel AT, and that of a battery RT is 2% below that of a diesel RT, and the gap widens as diesel prices rise and BET prices fall. FCTs, by contrast, remain more expensive to own than diesel trucks at the assumed hydrogen price of 20 AUD per kg in 2026, except for fuel cell RTs from about 2045.

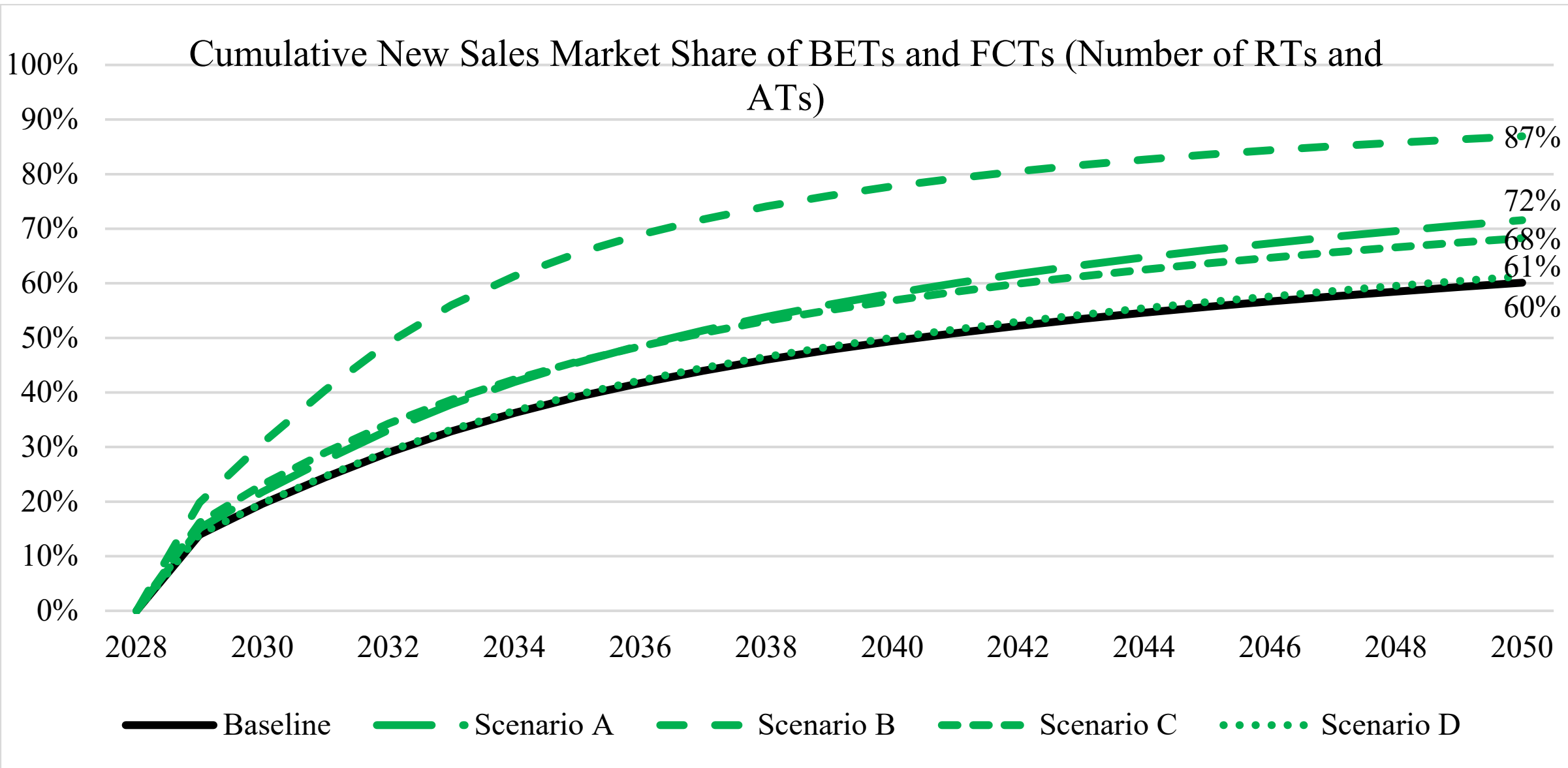

**Figure 5 | Cumulative share of BETs and FCTs in new RT and AT sales in energy-price scenarios A–D.**

Adoption nevertheless lags the TCO advantage. Battery ATs become as attractive as diesel ATs only in 2031, four years after they become available (Figure 6), and zero-emission RTs as attractive as diesel RTs in 2032. Two mechanisms explain this lag. First, the perceived adequacy of the charging network, which multiplies the attractiveness of BETs, builds up gradually, from 0.44 in 2030 to 0.79 in 2035 and 0.92 in 2040. Second, new vehicles replace the existing stock only gradually. Consequently, ZEHVs represent 62% of the RT and AT fleet and 70% of its freight task in 2050, and more than a third of the fleet still runs on diesel.

Figure 6 compares the relative attractiveness of diesel and battery ATs, which governs the allocation of new AT purchases. In the baseline, battery ATs match diesel ATs in early 2031, four years after market entry, because their attractiveness is multiplied by the gradually rising perceived adequacy of the charging network. Scenario B brings parity forward to 2028, Scenarios C and A to 2029, whereas Scenario D advances it by only two months. Zero-emission RTs (BET and FCT combined) reach parity with diesel RTs in 2032 in the baseline, 2030 in Scenario B, 2031 in Scenarios A and C and 2032 in Scenario D.

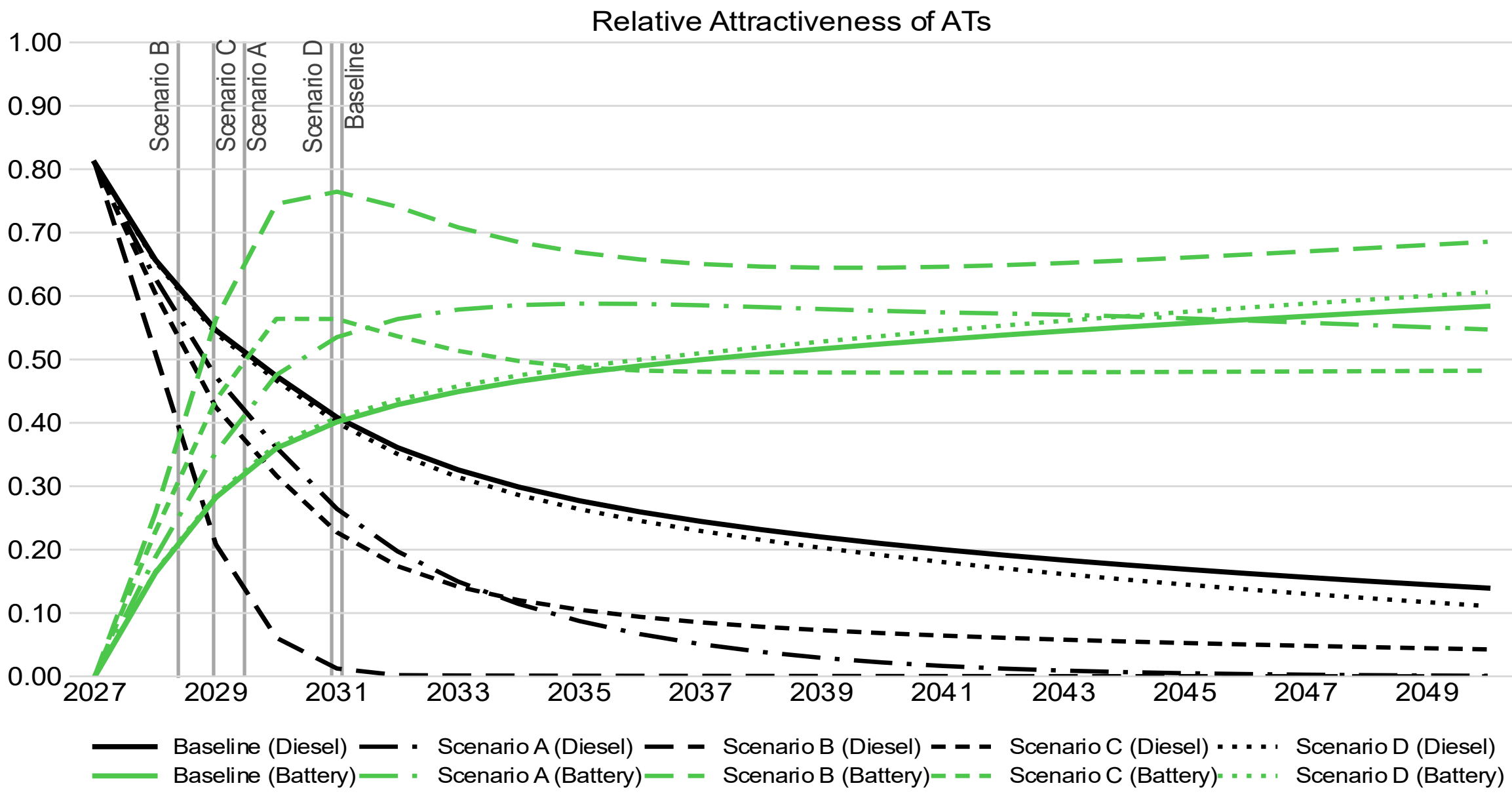

**Figure 6 | Relative attractiveness of diesel and battery electric ATs under the energy-price scenarios; vertical lines mark the year of parity.**

## Size, Timing and Persistence in Energy-Price Scenarios

Scenario B produces both the fastest and the largest shift (Figure 5). The short-term energy price shock raises the cumulative zero-emission share to 66% by 2035 and to 87% by 2050, and ZEHVs account for half of annual sales from 2030. Scenario A raises the cumulative share to 72% in 2050 and Scenario C, in which electricity and hydrogen become almost free, to 68%; ZEHVs reach half of annual sales in 2032 and 2031, respectively, compared with 2033 in the baseline. Scenario D, the realistic RUC benchmark, barely departs from the baseline (61% in 2050). The shares of the fleet and of the freight task follow the same ranking, with one notable exception. Although Scenario C adds less to the cumulative sales share than Scenario A, it raises the zero-emission share of the 2050 freight task to 85%, close to Scenario A (89%). Cheap electricity and hydrogen favour ATs, for which energy is about a third of OPEX, and shift freight from RTs to ATs, so the change is concentrated in the segment that carries most of the freight with relatively few vehicles.

These results do not support H2 in its original form. Scenario B outperforms Scenario A on every adoption indicator, but the two shocks are not of comparable magnitude. In Scenario B the diesel price remains about twice that of Scenario A in 2050. The comparison does, however, show the importance of timing. In Scenario B, battery ATs become as attractive as diesel ATs by 2028, about a year earlier than in Scenario A, and by 2035 the cumulative zero-emission share is 66% in Scenario B but 45% in Scenario A, so ZEHVs enter the fleet earlier and remain in service for their full lifetimes. Because each purchase decision rests on a discounted, multi-year TCO comparison, a large and early cost gap is locked into the fleet for 12–15 years, whereas the two-sided price shift in Scenario A builds a comparable cost gap only gradually. A strong early price signal therefore reduces cumulative emissions most, provided that the cost advantage it creates is not subsequently reversed.

### Push versus Pull, Operating versus Purchase Costs

All five policies (E-I) accelerate adoption relative to the baseline, but by markedly different amounts (see the supplementary file, Figure S1). The combined package (Scenario E) performs best; the cumulative zero-emission share reaches 44% by 2035 and 65% by 2050 (5.1 percentage points above the baseline), and ZEHVs account for half of annual sales from 2032, a year earlier than in the baseline. Among the single instruments, the OPEX subsidy (Scenario G) has the largest effect (62.0%, 1.9 percentage points above the baseline), followed by the carbon price (Scenario I; 61.5%), the CAPEX subsidy (Scenario F; 61.1%) and the RUC increase (Scenario H; 61.1%), which add 1.0–1.4 percentage points each.

In terms of total fleet market share, by 2050, ZEHVs account for 66.4%, 62.8%, 63.4%, 63.1% and 63.6% of the RT and AT fleet under Scenarios E, F, G, H and I, respectively, compared with 62.0% in the baseline, and for 78%, 71%, 74%, 72% and 73% of its freight task (70% in the baseline). The gains of the individual instruments sum to 5.3 percentage points, close to the gain of the combined package (5.1 percentage points), so at these levels the instruments are nearly additive: they act on different cost components, and none is large enough to exhaust the purchases that are most sensitive to cost.

Figure 7 shows that the policy scenarios bring the attractiveness parity of battery ATs forward by less than two years. Scenario E advances it by about 20 months and Scenario G by about 11 months, whereas Scenarios F, H and I advance it by only 2–7 months. For zero-emission RTs, parity moves from 2032 in the baseline to 2031 under Scenario E (about 11 months earlier) and by 1–4 months under the single instruments. Because battery trucks already reach parity in the early 2030s in the baseline, the policies change the timing of adoption less than its pace in the following years.

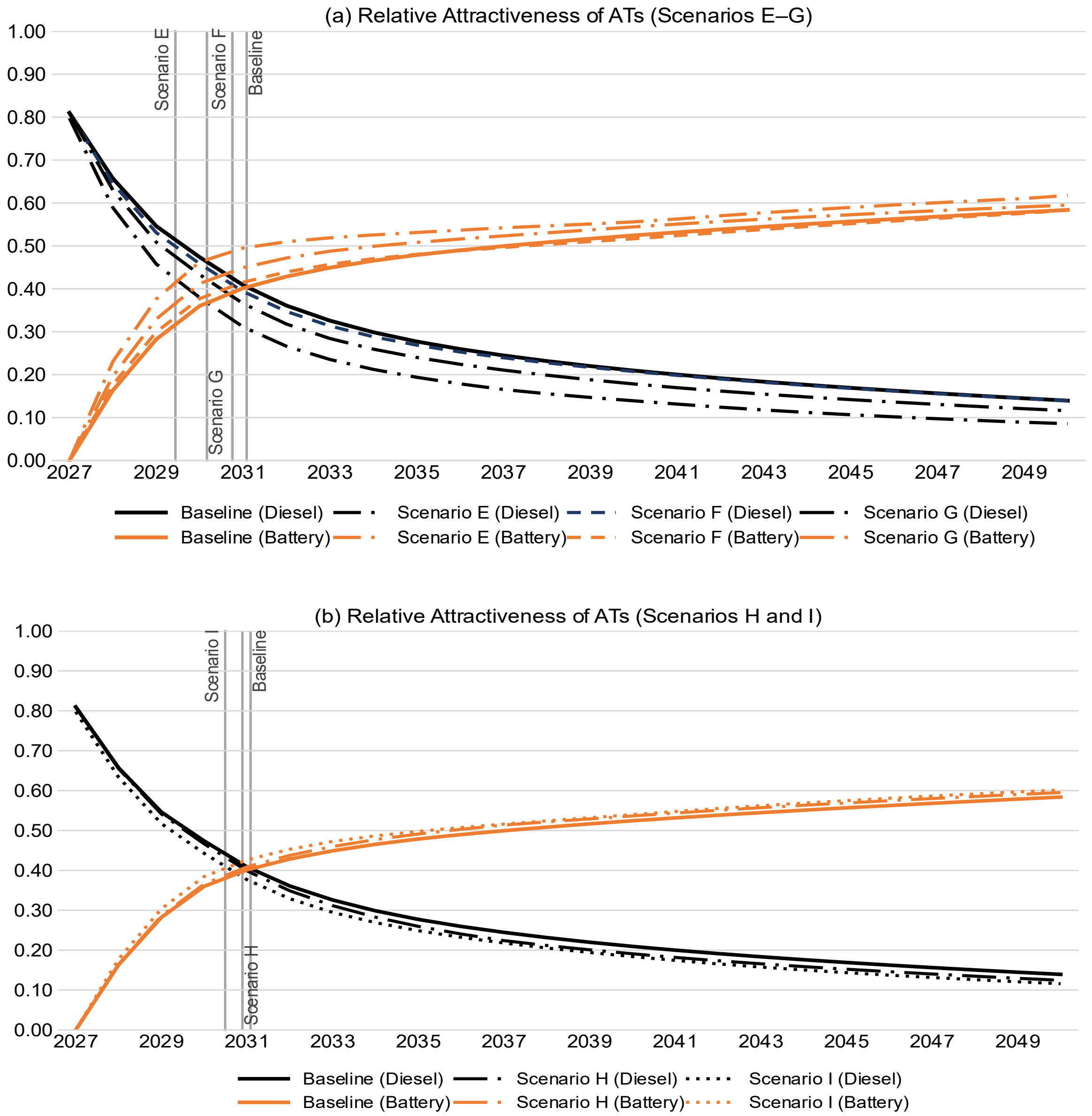

**Figure 7 | Relative attractiveness of diesel and battery electric ATs under the policy scenarios: (a) Scenarios E–G; (b) Scenarios H and I.**

## Elasticities

Figure 8 reports the dynamic elasticities implied by the energy-price scenarios, defined as the percentage change in cumulative new sales of a truck type relative to the baseline divided by the percentage change in the relevant price.

In Scenario D, the RUC is embedded in the pump price of diesel, as it is in Australia's fuel excise system, so Figure 8 (a-b) are both computed with respect to the diesel pump price. The elasticity of cumulative diesel-truck sales with respect to the diesel price is correctly signed but small, at −0.12 for ATs and −0.04 for RTs in 2029 and −0.10 and −0.07 by 2050 in Scenario B (Figure 8(a)), and growing in magnitude to −0.37 for ATs and −0.15 for RTs by 2050 in Scenario D (Figure 8(b)).

Demand for diesel trucks is therefore inelastic to fuel prices, which explains why even a tenfold price shock (Scenario B) shifts adoption only gradually; the elasticities are smaller in Scenario B than in Scenario D because the response saturates as the price change grows. They are up to three times larger for ATs than for RTs, because fuel accounts for about a third of the annual OPEX of a diesel AT but only about a tenth of that of a diesel RT, whose costs are dominated by driver wages. Zero-emission AT sales respond much more strongly to the price of electricity and hydrogen (Scenario C, Figure 8(c)). The elasticity is −0.7 in 2028 and declines in magnitude to −0.4 by 2050 as the price reduction approaches 100%. For RTs, it is small (−0.09 in 2028) and turns positive from 2029, not because cheaper energy deters zero-emission RT purchases, but because cheaper zero-emission ATs attract freight away from RTs, so that fewer RTs of all types are purchased.

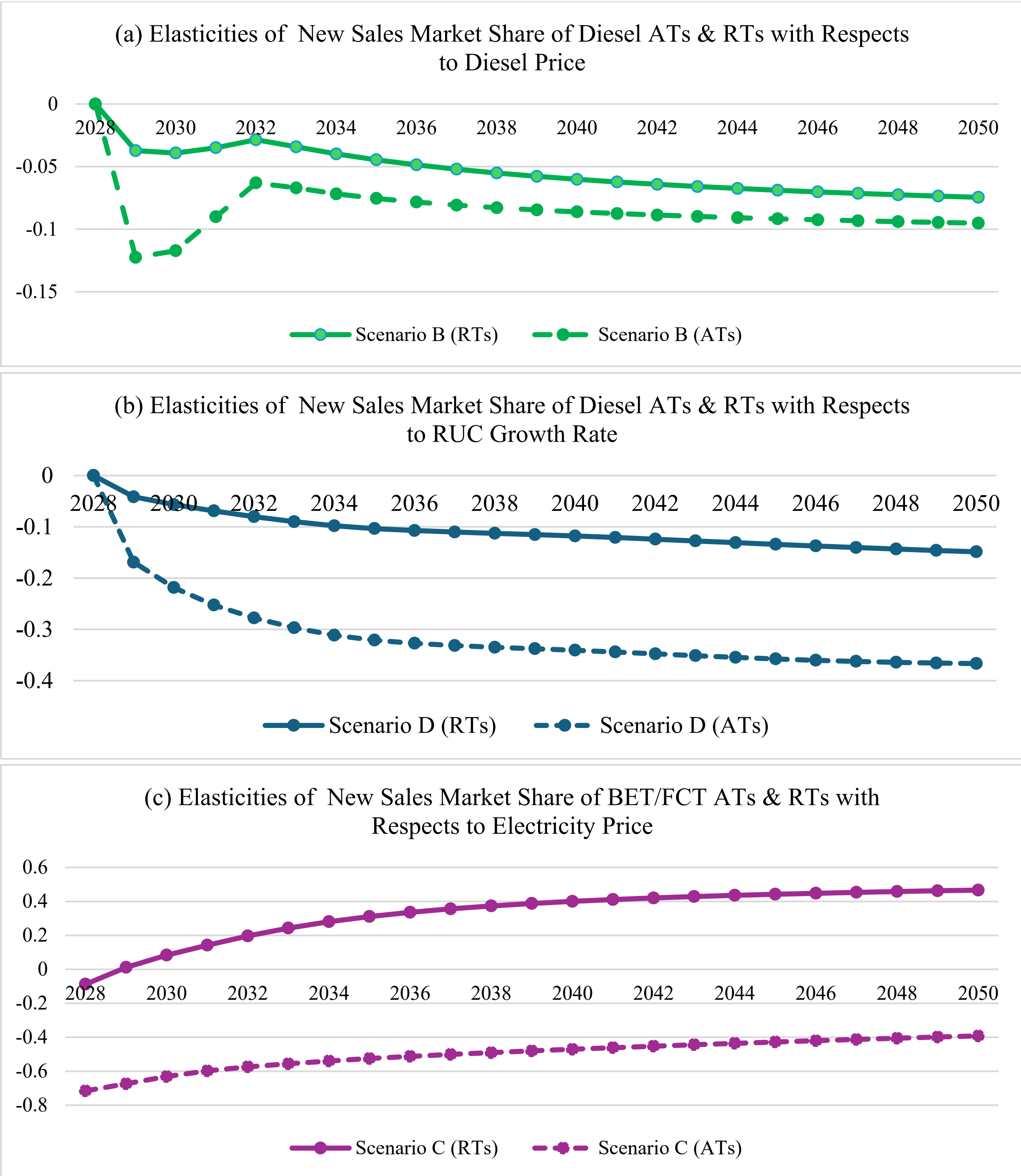

**Figure 8 | Elasticities of cumulative new sales under the energy-price scenarios: (a) diesel RTs and ATs with respect to the diesel price, Scenario B; (b) the same, Scenario D (RUC embedded in the diesel price); (c) BET and FCT RTs and ATs with respect to electricity and hydrogen prices, Scenario C.**

Figure 9 presents the elasticities with respect to CAPEX and OPEX cost reductions as a result of subsidies. The ranking of the single instruments reflects the size of the cost changes they produce. The elasticities in Figure 9 normalise for size. The elasticity of new ZEHV truck sales with respect to the CAPEX of BETs and FCTs (Scenario F) lies between −0.25 and −0.45 for ATs and between −0.17 and −0.36 for RTs (Figure 9(a)), whereas the elasticity with respect to their OPEX (Scenario G) is −2.6 for ATs in 2028, −2.1 in 2030 and still −1.2 in 2050 (Figure 9(b)).

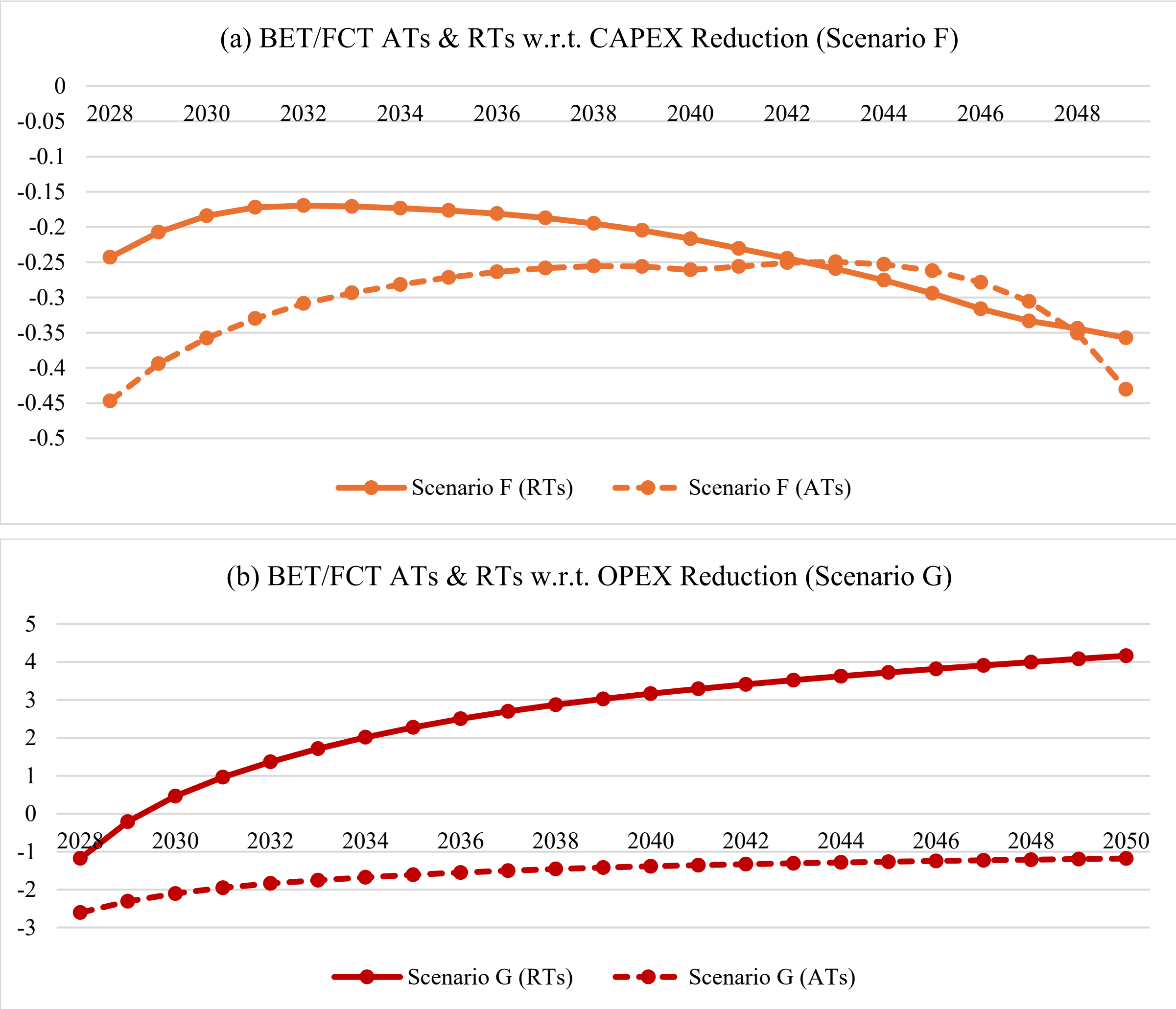

**Figure 9 | Elasticities of cumulative new BET and FCT sales with respect to (a) their CAPEX (Scenario F) and (b) their OPEX (Scenario G).**

Comparing the same years, the AT elasticity with respect to OPEX is about six times that with respect to CAPEX over 2028–2035 and five times in 2040, falling to about three by 2049, when the CAPEX subsidy is close to zero. Per percentage change in cost, OPEX instruments are therefore about five to six times as effective as CAPEX instruments for ATs while the subsidy is in place, consistent with H3. The OPEX subsidy has a modest overall effect only because it applies to energy and maintenance costs, which account for about a quarter of the annual OPEX of a battery AT and less than a tenth of that of a battery RT, so that a 20% subsidy lowers their OPEX by only about 5% and 1%, respectively (about 10% and 3% for fuel cell trucks). The CAPEX subsidy has a smaller effect still, although it covers 60% of the price premium, because it lowers the CAPEX of BETs by only 13–19% in 2027, shrinks with the premium and ends within 13–20 years, and because CAPEX accounts for only about 15% of the levelised TCO of a battery AT and 8% of that of a battery RT in 2030. For RTs, the OPEX elasticity is −1.2 in 2028 but turns positive from 2030 (Figure 9(b)), for the same reason as in Scenario C; the zero-emission share of new RT sales is higher in Scenario G than in the baseline throughout, but ATs attract part of the freight task away from RTs, so fewer RTs of all types are purchased.

Two further features of the results bear on hypotheses H3 and H4. First, in the levelised TCO (Eq. 1), the CAPEX is incurred once whereas OPEX recur over the vehicle's life, so CAPEX changes move the TCO comparison less than OPEX changes of the same percentage, and the purchase subsidy shrinks as the price premium of ZEHVs falls. This is consistent with previous studies showing that purchase grants have only marginal effects on HV TCO (Noll et al., 2022, Rout et al., 2022). Second, the effect of push measures depends on how far they raise the cost of diesel. Raising the RUC to the full excise (Scenario H) lifts the diesel pump price by only 8–9% and shifts adoption about half as much as the OPEX subsidy, although the elasticity of cumulative diesel-truck sales with respect to the diesel price under Scenario H (from −0.14 in 2029 to −0.65 in 2050 for ATs, and from −0.04 to −0.25 for RTs; computed as in Figure 8 but not shown) is of the same order as under Scenario D. The carbon price adds a similar amount to the diesel price (about 0.19 AUD per litre, or 8%, in 2027, rising to 0.62 AUD per litre, or 13%, in 2050) but applies in full from 2027, and therefore shifts adoption somewhat more than the RUC increase (1.4 versus 1.0 percentage points).

The response to the RUC is gradual rather than showing a threshold. Raising the RUC by 15%, 20% and 30% a year over 2027–2037 lifts it to 1.45, 2.38 and 6.33 AUD per litre in 2037 (compared with 0.69 AUD per litre in Scenario H and 0.42 AUD per litre in the baseline) and the diesel pump price by 33%, 62% and 187%. These increases raise the 2050 cumulative ZEHV share by 3.3, 5.7 and 12.4 percentage points (compared with 1.0 in Scenario H) and cumulative RUC revenue from RTs and ATs over 2027–2050 to 125, 185 and 393 billion AUD (48.7 billion AUD in the baseline). Adding ten percentage points through the RUC alone would therefore require an RUC of about 5 AUD per litre by 2037, about 15 times the 2025–26 rate of 32.4 cents per litre (the temporary reductions in 2026 are not modelled).

### Rigid versus Articulated Trucks

Assuming BET/FCT are available in the market for both RTs and ATs, the responses differ systematically between segments. In every scenario except F, the gain in the share of new BET/FCT sales is larger for ATs than for RTs. By 2050, Scenario E raises it by 9.0 percentage points (pp) for ATs (to 76%) but by 3.8 pp for RTs (to 62%); the RUC increase (Scenario H) raises it by 1.7 and 0.8 pp for ATs and RTs, and the OPEX subsidy (Scenario G) by 4.3 and 0.9 pp, respectively, whereas the CAPEX subsidy (Scenario F), which is larger relative to the price of RTs, raises it by 0.6 pp for ATs and 1.1 pp for RTs. The difference is bigger under the energy-price scenarios (Figure 10, see the supplementary file, Figure S3 for policy Scenarios E-I). Scenario A raises the share by 20 pp for ATs but 7 pp for RTs, and Scenario C by 18 pp for ATs but only 1.1 pp for RTs. Contrary to H5, ATs are therefore the more responsive segment in the model.

There are two potential explanations for this result. First, energy accounts for about a third of the OPEX of an AT but only about a tenth of that of an RT, so instruments that act on fuel or energy costs change the TCO of ATs far more. Second, because the segments compete for the same freight task, instruments that favour ZEHV ATs also shift freight from RTs to ATs. By 2050, the RT share of the freight task falls from 17.2% in the baseline to 15.9% in Scenario E, 10.9% in Scenario A and 8.9% in Scenario C, so fewer new RTs are purchased and the RT fleet renews more slowly, whereas push measures that raise diesel costs leave it almost unchanged (17.3% in Scenario H). This substitution is a feature of the single-pool allocation in the model and is likely to overstate the substitutability of rigid and articulated trucks, which largely serve different tasks. When the segment shares of new capacity are instead held at their 2027 values (see the Sensitivity Analysis section), Scenario A raises the ZEHV share of new RT sales by 14 pp rather than 7 pp, while its effect on ATs is almost unchanged (19 pp), so the substitution roughly halves the measured response of RTs but does not explain the larger response of ATs. Because ATs carry about four-fifths of the RT and AT freight task, their responsiveness drives the aggregate outcomes. This result is, however, conditional on the assumed commercial availability of BET/FCT ATs from 2027 and 2030, which is imposed exogenously in the model.

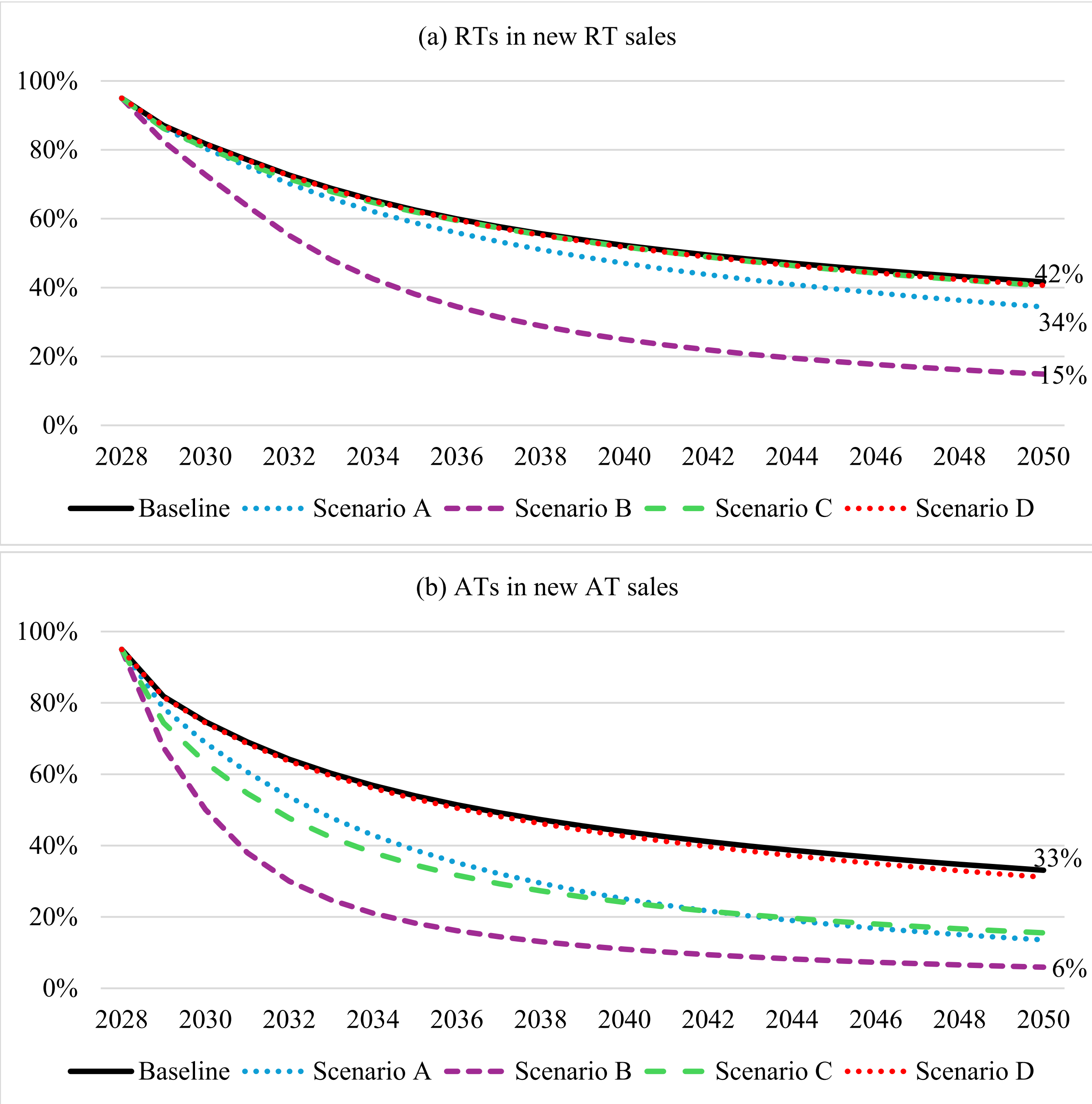

**Figure 10 | Cumulative share of diesel trucks in energy-price Scenarios A-D: (a) RTs in new RT sales; (b) ATs in new AT sales.**

## Infrastructure Perception and the Role of Hydrogen

With an adjustment time of five years, the perceived adequacy of the charging network that multiplies BET attractiveness rises from 0.44 in 2030 to 0.79 in 2035, 0.92 in 2040 and 0.99 in 2050, and the corresponding score for hydrogen refuelling from 0.62 in 2035 to 0.86 in 2040 and 0.98 in 2050. As adoption accelerates, however, physical charging capacity becomes a binding constraint. In the baseline, charging demand exceeds capacity only marginally, by at most about 1% over 2034–2038, but the shortfall grows with the pace of adoption. At its lowest, charging capacity covers 90% of charging demand in Scenario E, 87% in Scenario C, 84% in Scenario A and only 57% in Scenario B, in which demand exceeds capacity from 2029 to the end of the projection. The perceived adequacy of the charging network falls accordingly, to 0.75 in Scenario E, 0.71 in Scenario A and 0.52 in Scenario B in 2035, compared with 0.79 in the baseline. Hydrogen refuelling capacity exceeds demand by at least 16% in all scenarios. The infrastructure feedback therefore dampens adoption precisely in the scenarios in which adoption is fastest, and the speed at which operators come to trust the network remains a key uncertainty of the analysis.

Hydrogen nevertheless plays a role. FCTs account for 17–29% of cumulative new RT and AT sales by 2050 across scenarios (17% in the baseline) and for 25–43% of annual sales in 2050, even though, at the assumed hydrogen price of 20 AUD per kg, their levelised TCO remains 12–40% above that of BETs, because the allocation of purchases by relative attractiveness (Eq. 6) gives every available technology a share in proportion to its attractiveness. Selling hydrogen at the same price per unit as diesel at the pump from 2030 raises the FCT share of cumulative sales to 29%, largely at the expense of BETs (see the Sensitivity Analysis section). These shares should be read as conditional on the assumed hydrogen price and fuel-cell efficiency, and future studies can further fine-tune the cost assumptions related to the hydrogen FCTs.

## Diesel Use, Emissions and Fuel-Tax Revenue

Figure 11 presents the resulting diesel consumption of RTs and ATs in energy-price scenarios (A-D). In the baseline, diesel use falls from 7.8 billion litres in 2027 to 2.7 billion litres in 2050 (−66%) as ZEHVs gradually enter the fleet. Among the energy-price scenarios, diesel use in 2050 is 69% below the baseline under Scenario B, 55% below under Scenario A, 48% below under Scenario C and 5% below under Scenario D. The differences are less significant in the policy scenarios (see the supplementary file, Figure S2). The diesel consumption is 19.6% below the baseline under Scenario E, 11.0% below under Scenario G, 5.1% below under Scenario I, 3.4% below under Scenario H and 1.2% below under Scenario F. In cumulative terms over 2027–2050, Scenario B saves 32.2 billion litres of diesel relative to the baseline, equivalent to about 87 Mt of tailpipe $CO_2$ at 2.7 kg $CO_2$ per litre, followed by Scenarios C (58 Mt), A (52 Mt), E (24 Mt) and G (14 Mt), whereas the carbon price (Scenario I, 4.7 Mt), the CAPEX subsidy (Scenario F, 3.1 Mt), the RUC increase (Scenario H, 3.0 Mt) and the RUC benchmark (Scenario D, 2.9 Mt) save little. Scenario C saves more diesel than Scenario A despite its smaller gain in the cumulative sales share, because its effect is concentrated in ATs, which consume most of the diesel, and occurs earlier.

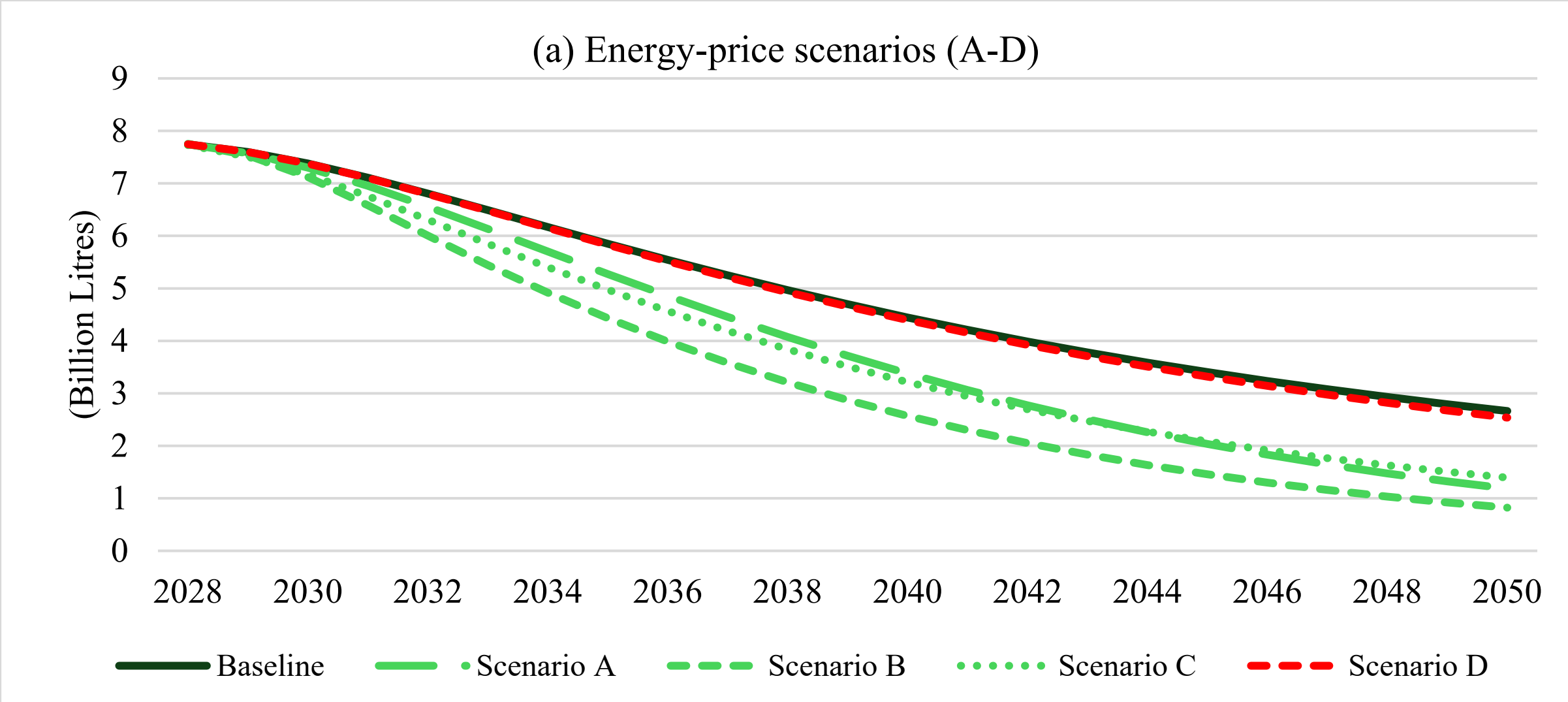

**Figure 11 | Diesel consumption of RTs and ATs (billion litres) in Scenarios A–D.**

Figure 12 shows the corresponding tailpipe $CO_2$ intensity of road freight, i.e. the total $CO_2$ emissions of LCVs, RTs and ATs per tonne-km of the freight task they carry. In the baseline, the intensity falls by 60%, from 0.167 kg $CO_2$ per tonne-km in 2027 to 0.067 kg in 2050. It falls to 0.021 kg in 2050 under Scenario B, 0.036 kg under Scenario A, 0.047 kg under Scenario C and 0.058 kg under Scenario E, whereas Scenarios D and F–I remain within 0.005 kg of the baseline. Even under the most effective policy scenario, road freight in 2050 therefore emits about a third as much $CO_2$ per tonne-km as in 2027, rather than approaching zero.

These avoided emissions are tailpipe (tank-to-wheel) emissions. Relative to the baseline, the additional electricity used by BETs over 2027–2050 ranges from about 1.5 TWh in Scenario F to 84 TWh in Scenario B (15 TWh in Scenario E), whereas Scenario C uses 23 TWh less electricity because cheap hydrogen shifts purchases towards FCTs; the additional hydrogen used by FCTs ranges from 0.1 Mt to 3.2 Mt (0.8 Mt in Scenario E). Each 0.1 kg $CO_2$ per kWh of average grid emissions intensity therefore offsets about 1.5 Mt (Scenario E) to 8 Mt (Scenario B) of the avoided tailpipe emissions, or 6–10% of them, and each 1 kg $CO_2$ per kg of hydrogen offsets about 0.8–3.2 Mt (3–6%). Because electricity-sector emissions are projected to fall by more than 60% between 2025 and 2030 as Australia moves towards 82% renewable generation (DCCEEW, 2025) , the upstream emissions of BETs are likely to offset only a small part of these gains, whereas hydrogen produced from unabated natural gas, at roughly 10 kg $CO_2$ per kg, would offset about a third of them in Scenarios B and E and more than half in Scenario C. Upstream emissions from diesel production, which are avoided together with tailpipe emissions, are not included and would partly compensate.

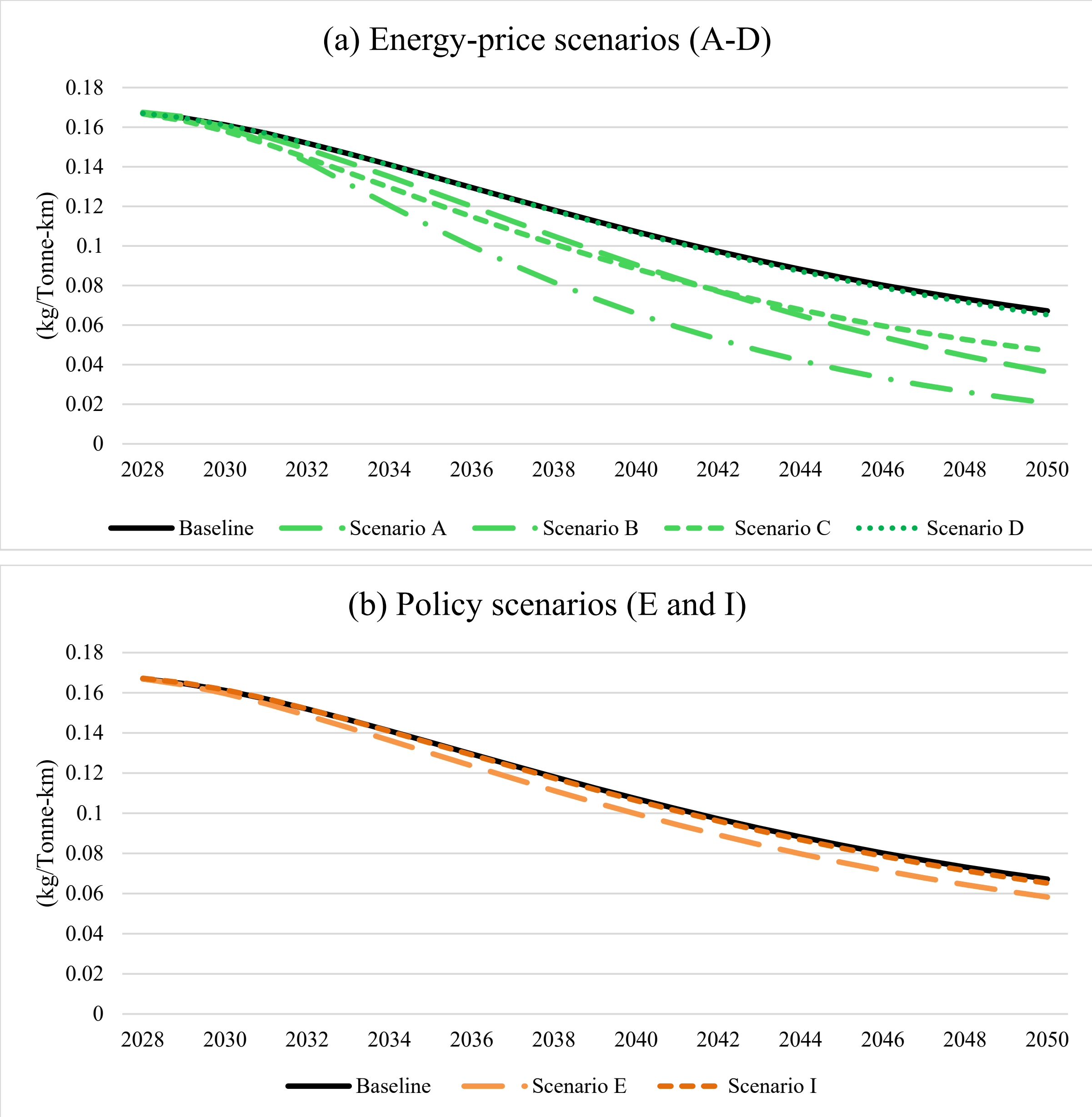

**Figure 12 | Total $CO_2$ emissions per tonne-km of road freight carried by LCVs, RTs and ATs (kg $CO_2$ per tonne-km): (a) Scenarios A–D; (b) Scenarios E–I.**

The scenarios also differ sharply in their fiscal implications (Figure 13). We measure fuel-tax revenue as the full fuel excise paid on the fuel used by LCVs plus the RUC component paid by RTs and ATs; carbon-price revenue is excluded. In the baseline, revenue rises from 6.7 billion AUD in 2027 to a peak of 7.4 billion AUD in 2033 and then declines to 6.1 billion AUD in 2050, giving cumulative revenue of 158 billion AUD over 2027–2050, of which 48.7 billion AUD is RUC paid by RTs and ATs. Because push measures tax a diesel fleet that shrinks only slowly, raising the RUC to the full excise in Scenario H lifts annual revenue to a peak of 8.6 billion AUD in 2037 and 7.0 billion AUD in 2050, and cumulative revenue to 180 billion AUD (+14%); the RUC paid by RTs and ATs rises by 45%, to 70.5 billion AUD. Scenario E generates 170 billion AUD and Scenario D 183 billion AUD. Pull measures and falling fossil-fuel demand, in contrast, erode the tax base where cumulative revenue falls to 156 billion AUD in Scenario F and 153 billion AUD in Scenario G, before the cost of the subsidies themselves is considered, and to 136, 138 and 111 billion AUD in Scenarios A, C and B, respectively. These results illustrate the fuel-tax erosion balancing loop: the more successful a policy is in displacing diesel, the faster it erodes the revenue from which infrastructure is funded. They also point to a fiscally robust design in which push-side revenue is recycled into pull-side support and infrastructure, as in Scenario E.

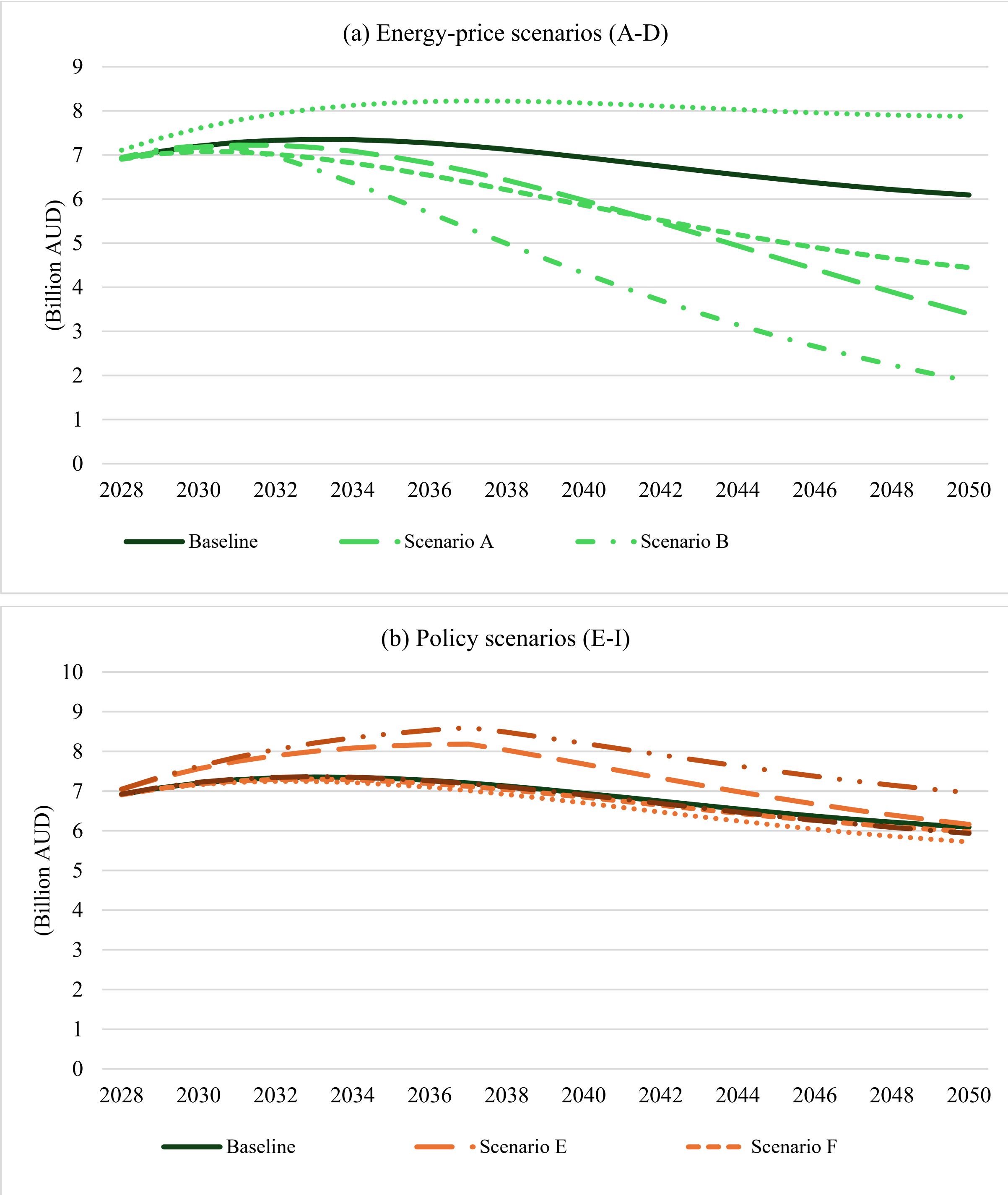

Figure 13 | Fuel-tax revenue: fuel excise paid by LCVs plus RUC paid by RTs and ATs (billion AUD per year): (a) Scenarios A–D; (b) Scenarios E–I.

## Sensitivity Analysis

Because several behavioural parameters cannot be estimated from historical data, the baseline and all nine scenarios were tested with the cost-sensitivity exponent $\gamma$ set to 2.5 and 7.5 instead of 4.73, with the adjustment time of perceived infrastructure adequacy $\tau_S$ set to 2 and 10 years instead of 5, and with the segment shares of new capacity held at their 2027 values (1.3% for LCVs, 17.7% for RTs and 81% for ATs) from 2027 onwards, so that fuels compete only within each segment (Table 6).

**Table 6: Cumulative ZEHV share of new RT and AT sales in 2050 (%) under alternative behavioural assumptions; in parentheses, the gain relative to the baseline under the same assumptions (percentage points)**

| Scenario | Baseline ($\gamma$ = 4.73, $\tau_S$ = 5 years) | Segment shares fixed | $\gamma$ = 2.5 | $\gamma$ = 7.5 | $\tau_S$ = 2 years | $\tau_S$ = 10 years |
|---|---|---|---|---|---|---|
| Baseline | 60.1 | 60.4 | 57.9 | 61.7 | 64.4 | 53.3 |
| Scenario A | 71.6 (+11.5) | 75.5 (+15.1) | 68.1 (+10.2) | 72.9 (+11.2) | 76.4 (+12.0) | 65.7 (+12.5) |
| Scenario B | 86.9 (+26.8) | 89.0 (+28.6) | 81.5 (+23.6) | 88.5 (+26.8) | 89.9 (+25.5) | 83.6 (+30.3) |
| Scenario C | 68.3 (+8.2) | 69.0 (+8.6) | 63.4 (+5.5) | 72.3 (+10.6) | 73.3 (+8.9) | 62.0 (+8.7) |
| Scenario D | 61.3 (+1.2) | 61.5 (+1.2) | 58.7 (+0.8) | 63.1 (+1.4) | 65.5 (+1.1) | 54.6 (+1.3) |
| Scenario E | 65.2 (+5.1) | 65.5 (+5.2) | 61.2 (+3.3) | 68.4 (+6.7) | 69.4 (+5.0) | 58.7 (+5.5) |
| Scenario F | 61.1 (+1.0) | 61.3 (+0.9) | 58.5 (+0.6) | 63.2 (+1.5) | 65.3 (+1.0) | 54.3 (+1.0) |
| Scenario G | 62.0 (+1.9) | 62.2 (+1.9) | 59.1 (+1.2) | 64.3 (+2.6) | 66.3 (+1.9) | 55.2 (+1.9) |
| Scenario H | 61.1 (+1.0) | 61.3 (+1.0) | 58.6 (+0.6) | 62.9 (+1.2) | 65.3 (+0.9) | 54.4 (+1.1) |
| Scenario I | 61.5 (+1.4) | 62.0 (+1.6) | 58.9 (+1.0) | 63.5 (+1.8) | 65.7 (+1.4) | 54.9 (+1.6) |

*Notes: Segment shares fixed: the shares of new freight capacity allocated to LCVs, RTs and ATs are held at their 2027 values (1.3%, 17.7% and 81%) from 2027, and fuels compete within each segment. $\gamma$ is the cost-sensitivity exponent of attractiveness (Eq. 4) and $\tau_S$ the adjustment time of perceived infrastructure adequacy (Eq. 5).*

The adjustment time is the most influential of these assumptions. With $\tau_S$ = 10 years, the 2050 cumulative ZEHV share of the baseline falls from 60.1% to 53.3% and battery-AT parity with diesel ATs is delayed to 2034, whereas $\tau_S$ = 2 years raises the share to 64.4% and brings parity forward to 2029. Varying $\gamma$ between 2.5 and 7.5 moves the baseline share by about two percentage points (57.9–61.7%), and fixing the segment shares by less than half a point. The gains of the scenarios relative to the baseline under the same assumptions are also robust. The ranking of the energy-price scenarios (B, A, C, D) is the same in all six variants, and among the policy scenarios the combined package (E) and the OPEX subsidy (G) rank first and second, followed by the carbon price (I), in all variants, whereas the CAPEX subsidy (F) and the RUC increase (H), whose effects differ by only 0.04 percentage points in the baseline, swap places in three of the six variants.

Varying the cost-sensitivity exponent $\gamma$ between 2.5 and 7.5 moves the share between 57.9% and 61.7%. Lower cost sensitivity ($\gamma$ = 2.5) reduces the gain of every scenario, by about a tenth for Scenarios A and B and by about a third or more for the others, and higher cost sensitivity ($\gamma$ = 7.5) raises the gains of Scenarios C–I by 20–50% while leaving those of A and B almost unchanged.

The adjustment time and the fixed segment shares change the gains by less than 15%, except that fixing the segment shares raises the gain of Scenario A by a third, because RTs can no longer lose freight to ATs.

A 10% payload penalty for battery RTs and ATs is the second most influential assumption. It lowers the share by 4.7 percentage points, to 55.4%, and delays the attractiveness parity of battery ATs with diesel ATs by about two years, to 2033.

Selling hydrogen at the same price per unit as diesel from 2030, which makes the fuel cost per tonne-km of FCTs about 80% lower than that of diesel trucks, raises it to 63.8%. However, the FCT share of cumulative sales rises from 17% to 29% largely at the expense of BETs, whose share falls from 43% to 35%.

The result is least sensitive to the discount rate used in the TCO, which lowers the share by only 0.6 and 1.5 percentage points at 10% and 15%, and to fixing the segment shares of new capacity at their 2027 values (+0.3 percentage points).

The projected transition therefore depends far more on how quickly operators come to trust the charging and refuelling network and on the payload of battery trucks than on financing costs or on how freight is allocated between segments. Figure 14 summarises the one-way sensitivity of the baseline result, the cumulative ZEHV share of new RT and AT sales over 2027–2050, to the main uncertain assumptions.

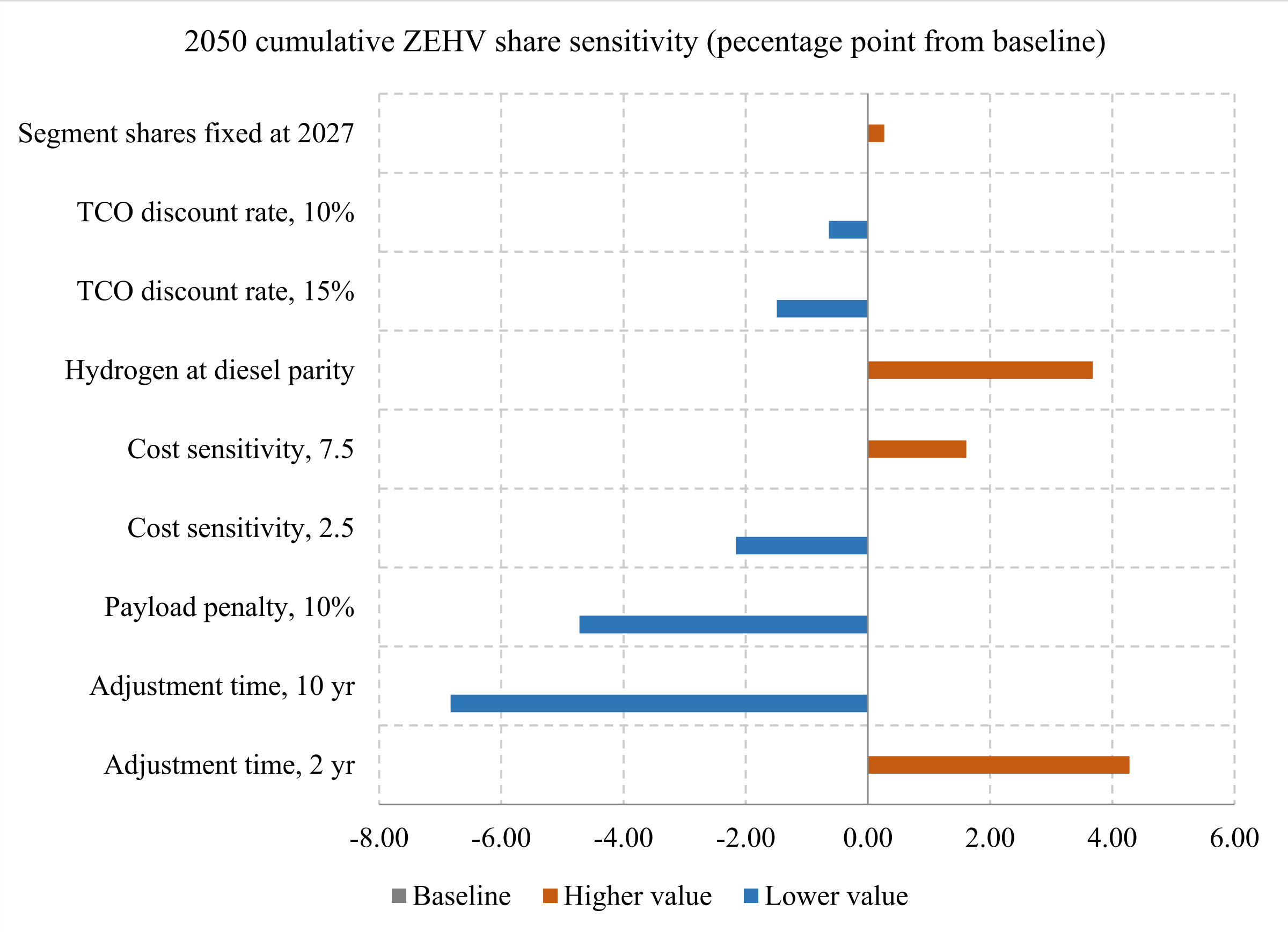

**Figure 14 | One-way sensitivity of the cumulative share of BETs and FCTs in new RT and AT sales over 2027–2050 in the baseline.** Bars show the increasing/decreasing change from the baseline value (60.1%) when one assumption is changed at a time.

## DISCUSSION AND CONCLUSIONS

This study develops a dynamic, SD-based decision support framework to examine the transition of the HV sector towards zero tailpipe emission technologies under a range of policy and market conditions. By integrating truck size choice, fleet turnover, energy prices, and policy interventions within a single

SD modelling structure, the analysis provides insights into both the pace and the structural drivers of freight decarbonisation. The hypotheses formulated earlier in the Introduction can now be assessed by this model.

H1 is supported. Under current settings, BET/FCT trucks account for 60% of cumulative new RT and AT sales over 2027–2050, and 62% of the fleet in 2050, and battery ATs do not become as attractive as diesel ATs until 2031, four years after market entry, even though their levelised TCO is below that of diesel ATs from market entry. More than a third of the RT and AT fleet, and 30% of its freight task, still rely on diesel in 2050.

H2 is not supported in its original form. The strong shorter-term shock (Scenario B) outperforms the persistent two-sided shift (Scenario A) on every indicator, although the two shocks differ in size. The comparison shows instead that the timing of a price signal governs cumulative $CO_2$ emissions.

H3 is supported. Per percentage change in cost, OPEX instruments are about five to six times as effective as CAPEX instruments for ATs, and the 20% OPEX subsidy is the most effective single policy instrument tested, although its absolute effect is modest because energy and maintenance costs are a small share of OPEX. Because OPEX still account for more than three-quarters of the TCO of BETs at a discount rate of 15%, this ranking does not depend on the discount rate.

H4 is supported only for large push measures. Raising the RUC to the full diesel excise (Scenario H) and the carbon price (Scenario I) shift adoption by 1.0 and 1.4 percentage points, less than the 20% OPEX subsidy (1.9 percentage points), because they raise the diesel price by less than 15%. Push measures outperform the pull measures tested only when they raise the cost of diesel substantially: an RUC that rises by 20% a year to 2.38 AUD per litre in 2037 adds 5.7 percentage points, about three times the effect of the OPEX subsidy. Push measures nevertheless generate the revenue needed to fund pull measures and infrastructure.

H5 is rejected. ATs respond more strongly than RTs in every scenario except the CAPEX subsidy (Scenario F), conditional on the assumed availability of alternative-fuelled AT models in the market. The difference persists when the segment shares of new capacity are fixed, so it does not arise only from the shift of freight between segments; a 10% payload penalty on battery trucks lowers the 2050 cumulative ZEHV share of new ATs by 4.3 percentage points, so the size of the AT response is an upper bound.

Lastly, H6 is supported. Charging capacity falls short of charging demand in every scenario, marginally in the baseline but by up to 43% in Scenario B, and in the scenarios with the fastest uptake (A, B, C and E) the resulting shortfall lowers the perceived adequacy of the charging network and dampens further adoption. The dampening is strongest under the extreme fossil-fuel shock, where capacity lags demand from 2029 to 2050. Hydrogen refuelling capacity, by contrast, keeps pace with demand in all scenarios.

The results show that the transition of Australia's HV fleet is inherently gradual and strongly path-dependent. Under current settings, BET and FCT account for 20% of cumulative new RT and AT sales by 2030 and 60% by 2050, and for 62% of the fleet in 2050, while diesel use falls by 66% but still produces about 7.2 Mt of tailpipe $CO_2$ in 2050. Adoption lags the TCO advantage of BETs by about four years. Although battery trucks are cheaper to own than diesel trucks when they become available in 2027, long vehicle lifetimes, slow fleet replacement and the slow build-up of confidence in the charging network mean that even large policy or price changes take a decade or more to register in the fleet composition. A target of zero-emission rigid and articulated fleet by 2050 is therefore out of reach without decisive action in the next few years. Apart from the extreme price shock, no scenario tested electrifies more than about three-quarters of the fleet by 2050. This partly reflects the model structure, in which every available technology retains a share of new sales in proportion to its attractiveness; as long as OEMs supply diesel trucks, some operators are likely to keep buying them.

The effectiveness of an instrument depends on both the target market and the size of the cost change it produces. Per percentage change in cost, the OPEX of ZEHVs is the most powerful lever. The elasticity of BET truck sales with respect to their OPEX is about five to six times that with respect to their purchase price, consistent with previous TCO studies showing that OPEX dominate HV ownership costs. A realistic subsidy that covers only energy and maintenance costs is the most effective single instrument tested, although its effect is modest (+1.9 percentage points in the 2050 cumulative share), because these costs are a small part of OPEX, which are dominated by driver wages. Raising the RUC to the full diesel excise by 2037 adds about one percentage point and raises cumulative RUC revenue from rigid and articulated trucks by 45%, and a carbon price of 70 AUD per tonne rising by 7 AUD a year, which raises the diesel price by 8–13%, adds 1.4 percentage points. Adding ten percentage points through the RUC alone would require it to rise to about 5 AUD per litre by 2037, some 15 times the 2025–26 rate of 32.4 cents per litre. A purchase subsidy covering 60% of the price premium of ZEHVs, as in Norway, adds only 1.0 percentage point, because CAPEX is a small part of the TCO and the premium, and with it the subsidy, shrinks over time. The combined push-and-pull package performs best among the policy scenarios, raising the cumulative BET/FCT share of new sales to 65% by 2050, with half of annual sales reached by 2032, reducing 2050 diesel use by 20% relative to the baseline and avoiding about 24 Mt of tailpipe $CO_2$ over 2027–2050. In our model, push measures at politically feasible levels therefore contribute less to adoption than to funding the pull measures and infrastructure that sustain it.

The fiscal feedback of the transition is among the most policy-relevant findings. Because push measures tax a diesel fleet that shrinks only slowly, raising the RUC to the full excise increases cumulative RUC revenue from RTs and ATs by 45%, from 48.7 to 70.5 billion AUD over 2027–2050, and a 20% annual increase over ten years would almost quadruple it (185 billion AUD), whereas pull measures and higher energy prices erode the fuel-tax base, by up to 30% under the extreme price shock (111 versus 158 billion AUD). The more successful the transition, the faster it erodes the revenue from which road and charging infrastructure are funded. Durable packages therefore need to pair push measures that raise revenue with pull measures and infrastructure that this revenue can fund, and future work should assess when and how a road user charge on BETs and FCTs could replace the eroding fuel-tax base without stalling adoption.

The analysis also reveals heterogeneity between vehicle segments that runs counter to our initial expectation. Articulated trucks, which carry about four-fifths of the rigid and articulated freight task and therefore dominate aggregate energy use and emissions, respond more strongly than rigid trucks in every scenario except the CAPEX subsidy. By 2050, the combined package raises the cumulative BET/FCT share of new ATs to 76% but that of new RTs to 62%. Energy accounts for about a third of the OPEX of articulated trucks but only a tenth of those of rigid trucks, whose costs are dominated by driver wages, so energy- and fuel-based instruments change the economics of articulated trucks far more. This result is conditional on BET/FCT articulated models becoming commercially available in Australia as assumed. For RTs, measures that accelerate fleet turnover, or that act on vehicle and labour-related costs, will be needed alongside energy-price instruments.

Energy-price dynamics shape transition pathways in two distinct ways. The size and persistence of a relative price change govern the long-run market share, and its timing governs cumulative emissions. A strong short-term fossil-fuel shock (Scenario B) raises the BET/FCT share of new sales fastest, brings battery-AT parity forward to 2028 and saves more than 30 billion litres of diesel over 2027–2050, because early adopters remain in the fleet for their full lifetimes. Very cheap electricity and hydrogen (Scenario C) act mainly on ATs and save slightly more diesel than a persistent two-sided price shift (Scenario A), while barely changing the composition of RT purchases. The current Strait of Hormuz crisis illustrates the first mechanism, where a sustained fuel-price shock can accelerate the transition, but only if the resulting cost advantage of BET/FCT is not subsequently eroded, for example by suspending road user charges. Conversely, the realistic RUC increase tested in Scenario D (6% per year) barely affects adoption, even though it more than doubles annual RUC revenue by 2050.

From a policy perspective, the results point to four priorities. First, governments should lower the OPEX of BET/FCT in a sustained and predictable way, for example through registration and road-user-charge exemptions, preferential electricity tariffs or support for depot-based solar generation and storage, because OPEX measures deliver the largest shift per unit of cost change; to be effective, such support must be large relative to total OPEX rather than confined to energy and maintenance costs. Second, push measures such as RUC increases should be retained, mainly because they generate the revenue needed to fund pull measures and infrastructure, offsetting the erosion of the fuel-excise base that a successful transition entails, although at politically feasible levels they accelerate adoption only slightly. Third, charging and hydrogen refuelling infrastructure must be planned ahead of demand and made visible to operators. In the model, adoption is delayed both by the time operators take to trust that the network is adequate and, in the scenarios with the fastest uptake, by charging capacity that lags behind demand, so early, well-signposted investment, together with flexible solutions such as battery swapping and mobile charging units at freight hubs and along key corridors (Section 1), can shorten this lag. Fourth, although not modelled in this study, policies that accelerate fleet turnover, such as scrappage schemes or BET/FCT sales targets, together with measures that secure the supply of BET/FCT articulated models suited to Australian operating conditions, are likely to be needed to overcome the inertia associated with long vehicle lifetimes.

Overall, effective decarbonisation of the HV sector requires coordinated, long-term policy packages that combine push measures, which accelerate adoption and fund the transition, with pull measures and infrastructure that lower the OPEX of BET/FCT and build operators' confidence, implemented early and sustained over time, and that explicitly account for fleet turnover, infrastructure perceptions and segment-specific dynamics. Sustaining the transition in the articulated segment, which dominates the freight task, will depend on vehicle availability and charging capacity keeping pace with demand, whereas accelerating the transition of rigid trucks will depend on measures that speed up fleet renewal.

This study has several limitations. First, the model is calibrated specifically to the Australian freight sector, using country-specific cost, tax, and regulatory parameters, so the results are not directly transferable to jurisdictions with different vehicle standards, energy markets, or policy settings. Second, each vehicle segment (RTs and ATs) is treated as internally homogeneous, aggregating a wide range of operator types, duty cycles, and financing positions into a single representative decision-maker per segment and fuel type; this abstracts away heterogeneity among freight operators, including the distinction between large fleet owners and the small, capital-constrained operators that dominate the Australian trucking industry. Third, the future commercial availability of battery electric and fuel cell articulated trucks is imposed as a pre-defined exogenous constraint rather than endogenously determined by global OEM production and allocation decisions; because Australia is a comparatively small, right-hand-drive market dependent on vehicle models designed primarily for Europe and China, the large modelled uptake in the articulated segment is conditional on this assumption holding, and would not materialise to the same extent if OEMs delay bringing suitable models to market. Fourth, vehicle costs, and behavioural parameters such as the cost-sensitivity exponent and the five-year adjustment time of perceived charging adequacy are represented as smooth, deterministic trends or fixed constants (including a single 6% discount rate) rather than as stochastic processes with associated uncertainty ranges, likely understating the true variability of outcomes.

In addition, the annual freight task per vehicle is the same for all fuels of a segment in the baseline, so the payload penalty and charging dwell time of battery trucks and the cost of depot chargers are not represented; all three favour battery trucks. With a 10% lower freight task for battery RTs and ATs, battery ATs keep a lower levelised TCO over their lifetime when they enter the market, but battery RTs reach TCO parity with diesel RTs only in about 2043; the attractiveness parity of battery ATs is delayed by about two years, and the 2050 cumulative ZEHV share falls by 4.7 percentage points, so the strong response of battery trucks, and of ATs in particular (H5), should be read as an upper bound. Fifth, the model is calibrated against a comparatively short historical window (2010–2023) and projected 25 years forward, a substantial extrapolation over which structural discontinuities in technology, trade, and policy are difficult to anticipate.

Sixth, emissions are estimated on a tailpipe (tank-to-wheel) basis using current emission factors, without a full lifecycle or well-to-wheel accounting (an indicative adjustment is given in Section 5) that would capture changes in the electricity generation mix or upstream hydrogen production pathway over the simulation period, and the infrastructure sub-model captures station-level profitability and lets charging capacity grow in response to demand without a binding maximum build rate, but does not explicitly represent the underlying physical constraints, such as electricity grid capacity or planning and approval timelines that could further slow real-world rollout.

A further limitation concerns the behavioural parameters of the attractiveness function. The cost-sensitivity exponent ($\gamma = 4.73$) and the segment constants were estimated from the behaviour of the diesel fleet over 2010–2023, when ZEHV sales in Australia were negligible, so their application to the choice between zero-emission and diesel trucks is an extrapolation. A lower value of $\gamma$ would slow, and a higher $\gamma$ accelerate, both the projected adoption and its response to policy. Varying $\gamma$ between 2.5 and 7.5 changes the 2050 cumulative ZEHV share of the baseline by about two percentage points and leaves the ranking of scenarios unchanged, apart from Scenarios F and H, which are virtually tied (Table 6); an out-of-sample test such as back-casting the observed uptake of ZEHVs in Norway or Sweden from their TCO would further bound this extrapolation. Because all truck types compete for a single pool of freight demand, the model also allows freight to shift between rigid and articulated trucks in response to their relative costs, which is likely to overstate the substitutability of segments that largely serve different tasks. Holding the segment shares of new capacity at their 2027 values, so that fuels compete only within each segment, changes the baseline result by less than half a percentage point but doubles the response of RTs in Scenario A; a nested structure, in which the freight task is allocated to segments by task type, would be a more complete solution. The projected role of hydrogen is also sensitive to the assumed hydrogen price and fuel-cell efficiency. Lastly, the five-year adjustment time of perceived infrastructure adequacy, which governs much of the lag between cost parity and adoption, is an assumption rather than an estimate, and it is the most influential parameter tested: adjustment times of 2 and 10 years give 2050 cumulative ZEHV shares of 64% and 53% in the baseline; formulations in which operators' willingness to consider zero-emission trucks depends on the installed base and on network coverage (Struben and Sterman, 2008) would make this feedback endogenous. Future work should collect revealed- or stated-preference data from Australian freight operators, for example through discrete choice experiments, to estimate the cost sensitivity of zero-emission truck purchase decisions and the influence of infrastructure perceptions.

Future research could extend this work by incorporating behavioural heterogeneity among freight operators (e.g. ancillary shippers versus for-hire carriers), and more detailed representations of vehicle sizes (e.g. light, medium, and heavy rigid), freight tasks (e.g. separation of first-last mile short-haul versus long-haul freight task), commodity types (e.g. containerised, bulk non-perishable, and perishable cargo), and regions. Further details on the supply-side constraints, such as vehicle availability and subsidies for charging infrastructure construction could also yield more realistic results. Such extensions would further enhance the ability of the presented SD model to inform robust and adaptive policy design in the context of freight decarbonisation.

## Appendix A: Total Cost of Ownership Studies

Economic feasibility is a crucial consideration for HV operators when purchasing and operating HVs, encompassing various factors captured in the concept of Total Cost of Ownership (TCO). TCO calculations look beyond the vehicle price or so-called capital expenditures (CAPEX) and include other major operational costs (OPEX) that could affect the purchase decision of various HV powertrains. As pointed out by Noll et al. (2022), using TCO is more appropriate to quantify cost-competitiveness compared to CAPEX for HVs because 75% of total costs are operational due to high driving distances. Furthermore, TCO is an easier tool for operators to track all HV costs. TCO calculations are iterative processes that rely on real-world experiences to generate a comprehensive list of assumptions used in the calculation. These assumptions are refined to capture the nuances of specific use cases, such as heavy trucks travelling 60,000 km per year, with each use case having distinct assumptions. Hence, it is crucial to understand their assumptions and parameters before comparing the obtained values of TCOs for various powertrains. Four key cost components are commonly addressed in non-scientific industry TCO calculations: CAPEX, fuel/energy costs, operating and maintenance costs, and infrastructure costs.

A review of the academic literature provides insights into the components and assumptions underlying TCO estimation. Table A1 summarises the TCO components and parameters considered across four comprehensive studies. It should be noted that these studies represent only a subset of the literature that has assessed the economic feasibility of BETs and/or FCTs using the TCO approach (e.g. see (Liu et al., 2021, Breuer et al., 2021, Nykvist and Olsson, 2021, Gnann et al., 2017, Li and Kimura, 2021, Li and Taghizadeh-Hesary, 2022, Ribberink et al., 2021)).

Noll et al. (2022) suggest that the competitiveness of ZEHVs (i.e. BET and FCTs) depends on each country and its incentive policies. Notably, they found that the TCO of BETs is relatively competitive with ICE trucks in progressive countries such as Switzerland, Germany, Sweden, and Norway. This is due to no-toll charges for ZEHVs in Switzerland and Germany, and a combination of using cheap electricity, high renewable energy penetration and high fossil-fuel costs in Norway and Sweden. Whereas other countries, such as Italy and Poland, with ineffective incentives, exhibit higher TCO for such vehicles, thus making BETs uncompetitive. According to their estimation, the TCO of ZEHVs comprises 75% OPEX and recommends policymakers target OPEX parameters instead of introducing CAPEX subsidies since the impact of CAPEX seems less effective for HVs. Possible reasons include lower fuel costs, operation and maintenance costs and the introduction of government incentives that favour ZEHVs. Rout et al. (2022) examined the implications of introducing a fossil fuel tax and BET/FCT purchase grant and concluded that purchase grants only have a marginal effect on TCO, while a fossil fuel tax can significantly increase the TCO of diesel trucks. Interestingly, purchase grants only reduce TCO by around 3-4%, which also verifies the finding from Noll et al. (2022) that operational parameters are more dominant for TCO.

Table A2 summarises the TCO components for BETs and FCTs that were considered in 9 recently published industry reports. While infrastructure costs have received less attention in TCO calculations, certain reports have included the expenses associated with building hydrogen refuelling stations and pipelines necessary to support the fleet. Although these costs are not directly related to the vehicles themselves, understanding the complete cost of ownership is advantageous for large corporations, which are the primary target of many TCO reports.

**Table A1 | TCO components considered in previous studies.**

| Study | Noll et al. (2022) | Rout et al. (2022) | Speth, Kappler et al. (2022) | Burke and Sinha (2020) |
|---|---|---|---|---|
| Study Region | Europe (10 Countries) | Europe (UK) | Europe (Germany) | North America (California, USA) |
| CAPEX | | | | |
| Energy Storage Costs | * | * | * | * |
| Powertrain Costs | * | * | * | * |
| Tractor (rest of the truck) | * | * | * | * |
| Overheads | | * | | |
| CAPEX Subsidies | * | * | * | |
| Depreciation | * | * | | * |
| Scrappage Value | * | * | | * |
| OPEX | | | | |
| Tolls | * | | * | |
| Fuel Costs Per Unit of Energy | * | * | * | * |
| Driver Wages | * | | | |
| Fuel Consumption | * | * | * | * |
| Operation and Maintenance Costs | * | * | * | * |
| Insurance | * | * | * | |
| Infrastructure Costs for BETs (NB: all other in the fuel cost) | * | | | |
| Vehicle Tax (e.g. vehicle excise duty, road user levy) | | * | * | |
| TCO Parameter Considerations | | | | |
| Truck Weight | * | | | * |
| Load Assumptions | * | | | |
| Power | * | | * | * |
| Energy on Board | * | | * | * |
| Operating Load % | * | | | |
| Lifetime (years) | * | * | * | * |
| Annual KM Travelled | * | | * | * |
| Tank to Wheel Efficiencies | * | | | * |
| Fuel Consumption | * | * | * | * |
| Daily Distance Travelled | * | * | * | * |

*Overview of comprehensive TCO modelling approaches for ZEHVs in the academic literature. The table illustrates the diversity of cost structures and operational assumptions adopted across previous studies, including energy storage systems, infrastructure allocation, fuel consumption, operating load, annual travel distance, and vehicle lifetime. These differences significantly influence the comparability and transferability of reported TCO estimates.*

**Table A2 | TCO components in industry reports.**

| Region | Study | CAPEX | Energy or Fuel | Insurance | Maintenance | Tax | Resale | Vehicle Type | VKT | Fuel cost |
|---|---|---|---|---|---|---|---|---|---|---|
| Australasia | QTLC (2022) | * | * | | * | | | Heavy rigid urban delivery, articulated line-haul | 40,000 km/yr heavy rigid, 250000 articulated | Not mentioned |
| | Advisian (2021) | * | * | | * | * | * | Light vehicles, line haul heavy duty, return to base heavy duty, mining vehicles, forklifts | Network upgrades included in vehicle price, assumed salvage values & static VKTs | Grey, blue, green hydrogen & diesel prices |
| Europe | (Basma et al., 2022a) | * | * | | | | | Light rigid trucks for delivery, 40-60 km per day, BET only | 15,000 km/yr | Electricity price |
| | (Basma et al., 2022b) | * | * | | * | * | * | Tractor-trailer FCT only | 158,000km/yr | Considered |
| | Earl et al. (2018) | * | * | * | * | | | BET only | 160, 480, 800km per service | Not mentioned |
| | Ruf et al. (2020) | * | * | * | * | * | * | Tractor & two rigid trucks | 140,000 km, 95,000 km, 60,000 km/yr | Considered |
| | Unterlohner (2020) | * | * | | * | * | | Regional delivery and long-distance vehicles | 400 km range, 1,200 km range. | Considered |
| North America | Hunter et al. (2021) | * | | | * | * | | Three long-haul trucks & one parcel delivery | Vehicles improve over time, different VKTs considered | Calculated breakeven for fuel types |
| Global | Deloitte (2020) | * | * | * | * | | | Long-haul heavy rigid trucks | Not mentioned | Hydrogen price over time |

*Comparison of TCO components and assumptions across industry reports and regions. The table highlights variations in the total cost of ownership (TCO) methodologies adopted in previous studies, including differences in assumed vehicle kilometres travelled (VKT), truck classes, and operational characteristics. Regional differences in fuel and energy costs, infrastructure availability, carbon pricing mechanisms, subsidies, and policy incentives also contribute to substantial variation in reported TCO estimates.*

## Appendix B: Model Parameters

**Table B1| Annual Growth Ratio**

| | Vehicle Purchase Costs | Fuel/Energy Price | RUC | Maintenance Costs |
|---|---|---|---|---|
| **Diesel trucks** | 0.54% | 3.05%* | 2.50%** | 0.12% |
| **BETs** | -1.34% | -0.89% | – | – |
| **FCTs** | -1.35% | -0.59% | – | -0.71% |

*Summary of exogenous growth rate assumptions for vehicle purchase costs, fuel/energy prices, RUC, and maintenance costs used in the modelling framework, differentiated by drivetrain technology. Adopted from (Burke and Sinha, 2020, Noll et al., 2022, Rout et al., 2022, Speth et al., 2022, Transport and Infrastructure Council, 2016) *Long-run trend of 4.11% per year scaled by the calibration factor of 0.742; the RUC component of the pump price grows separately. **6% per year over 2023–2026.*

**Table B2 | Average Fuel Consumption Price and Rates and Estimated Costs (AUD) in 2025**

| | Fuel | Energy Price (per litre/Kwh/Kg) | Average Fuel Consumption (litre/Kwh/Kg per tonne-km) | Upfront Purchase Cost | Stamp Duty | Maintenance Costs (per tonne-km) | Insurance Costs (per tonne-km) | Annual Registration Costs |
|---|---|---|---|---|---|---|---|---|
| LCV | Diesel | $2.13 | 0.628 | $35K | 3% of the purchase cost | 0.0134 | 0.05 | $1K |
| | Battery | $0.45 | 2.072 | $60K | | 0.0107 | 0.06 | |
| | H2 | $20 | 0.126 | $100K | | 0.0120 | 0.09 | |
| RT | Diesel | $2.13 | 0.076 | $100K | | 0.0061 | 0.005 | $3K |
| | Battery | $0.45 | 0.250 | $150K | | 0.0028 | 0.007 | |
| | H2 | $20 | 0.015 | $200K | | 0.0034 | 0.011 | |
| AT | Diesel | $2.13 | 0.023 | $300K | | 0.0102 | 0.001 | $6K |
| | Battery | $0.45 | 0.078 | $400K | | 0.0034 | 0.003 | |
| | H2 | $20 | 0.005 | $500K | | 0.0036 | 0.003 | |

*Assumed average fuel consumption, fuel prices, and resulting cost estimates (AUD) used in the TCO calculations for the year 2025; Maintenance costs: adopted from (Burke and Sinha, 2020, Noll et al., 2022, Rout et al., 2022, Speth et al., 2022, Transport and Infrastructure Council, 2016)); Purchase costs: adopted from (Burke and Sinha, 2020, Noll et al., 2022, Rout et al., 2022, Speth et al., 2022, Transport and Infrastructure Council, 2016)); Insurance costs and stamp duty: adopted from (Freight Metrics, 2025); Annual registration costs: adopted from (National Transport Commission, 2025).*

**Table B3| Summary of charging infrastructure model inputs**

| | Electric charging | Hydrogen refuelling |
|---|---|---|
| CAPEX of charging stations | 5 million AUD for 1 MWh | 10 million AUD for 1 tonne |
| OPEX of charging stations (% of revenue) | 22.5% | 37.5% |
| Station capacity | 1 MWh | 1 tonne per day |
| Maximum effective operational time | 24 * 365 hours | 365 days |
| Energy capacity factor | 0.89 | 0.92 |
| Energy distribution costs (% of energy price) | 15% | |
| Maximum lifetime | 25 years | |
| Discount rate | 6% | |
| Construction time | 2 years | |
| The percentage of fuel excise tax income if the policy is enabled | 20% | |
| Subsidy (% of CAPEX) | 20% | |
| Salvage rate (% of CAPEX) | 10% | |

*This table presents the key costs and parameters used in the charging infrastructure model, adopted from stakeholder consultations and industry reports (CSIRO and GHD, 2023).*

## Appendix C: Key Indicators in Scenarios

Table C1 summarises the key indicators for all scenarios. Three patterns stand out. First, even the most effective scenarios leave a substantial diesel fleet in 2050. This is, however, based on an assumption that the diesel, BET and FCT models are available for both RTs and ATs until 2050 and the fuel segments are still competing based on their relative attractiveness. Apart from the extreme price shock of Scenario B, ZEHVs make up at most about three-quarters of the RT and AT fleet by 2050, although they account for about 90% or more of annual sales in Scenarios A, B and C by then. Second, the ranking of scenarios differs between indicators. Among the energy-price scenarios, Scenario C adds less than Scenario A to the new sales share but saves more diesel, because it acts mainly on ATs. Among the policy scenarios, the OPEX subsidy (G) leads the single instruments on adoption and diesel savings, whereas the RUC increase (H) leads on fuel-tax revenue. Third, the push measures that raise substantial additional tax revenue on their own (Scenarios D and H) shift adoption only slightly, whereas the pull measures shift adoption but erode tax revenue; only the combined package (Scenario E) achieves both. This suggests that a road user charge for BETs and FCTs may eventually be needed to sustain revenue, although its timing and its effect on adoption were not tested here.

**Table C1: Key indicators by scenario for rigid and articulated trucks**

| Indicator | Base line | A | B | C | D | E | F | G | H | I |
|---|---|---|---|---|---|---|---|---|---|---|
| ZEHV share of new RT and AT sales since 2027, by 2035 (%) | 39.2 | 45.5 | 65.5 | 45.6 | 39.6 | 44.0 | 40.7 | 41.0 | 39.7 | 40.3 |
| ZEHV share of new RT and AT sales since 2027, by 2050 (%) | 60.1 | 71.6 | 86.9 | 68.3 | 61.3 | 65.2 | 61.1 | 62.0 | 61.1 | 61.5 |
| of which FCTs, 2050 (pp) | 16.8 | 25.3 | 29.4 | 26.8 | 17.2 | 20.2 | 17.9 | 18.2 | 17.2 | 17.4 |
| ZEHV share of new RT sales, 2050 (%) | 58.4 | 65.6 | 85.1 | 59.4 | 59.3 | 62.2 | 59.5 | 59.3 | 59.2 | 59.4 |
| ZEHV share of new AT sales, 2050 (%) | 66.9 | 86.4 | 94.1 | 84.5 | 68.9 | 75.9 | 67.5 | 71.2 | 68.6 | 69.9 |
| ZEHV share of annual new RT and AT sales, 2050 (%) | 77.3 | 98.5 | 99.6 | 87.8 | 79.7 | 82.5 | 77.5 | 79.3 | 78.6 | 79.3 |
| Year ZEHVs first exceed 50% of annual new RT and AT sales | 2033 | 2032 | 2030 | 2031 | 2033 | 2032 | 2033 | 2032 | 2033 | 2033 |
| ZEHV share of RT and AT fleet (number), 2050 (%) | 62.0 | 73.7 | 86.5 | 67.2 | 63.5 | 66.4 | 62.8 | 63.4 | 63.1 | 63.6 |
| ZEHV share of RT and AT freight task (tonne-km), 2050 (%) | 70.3 | 89.1 | 92.8 | 85.3 | 72.5 | 77.6 | 70.6 | 73.8 | 71.9 | 72.8 |
| Attractiveness parity, battery vs diesel ATs (year) | 2031 | 2029 | 2028 | 2029 | 2031 | 2029 | 2031 | 2030 | 2031 | 2030 |
| Attractiveness parity, ZEHV vs diesel RTs (year) | 2032 | 2031 | 2030 | 2031 | 2032 | 2031 | 2032 | 2032 | 2032 | 2032 |
| Diesel use by RTs and ATs, 2050 (billion L) | 2.66 | 1.19 | 0.82 | 1.39 | 2.54 | 2.14 | 2.63 | 2.37 | 2.57 | 2.53 |
| Cumulative diesel saving vs baseline, 2027–2050 (billion L) | – | 19.1 | 32.2 | 21.3 | 1.1 | 8.9 | 1.1 | 5.1 | 1.1 | 1.7 |
| Avoided tailpipe $CO_2$ vs baseline, 2027–2050 (Mt) | – | 51.5 | 86.9 | 57.6 | 2.9 | 24.0 | 3.1 | 13.7 | 3.0 | 4.7 |
| $CO_2$ intensity of road freight, 2050 (kg per tonne-km) | 0.067 | 0.036 | 0.021 | 0.047 | 0.065 | 0.058 | 0.066 | 0.063 | 0.066 | 0.065 |
| Fuel-tax revenue, 2050 (billion AUD) | 6.1 | 3.4 | 1.9 | 4.4 | 7.9 | 6.2 | 6.0 | 5.7 | 7.0 | 5.9 |

| Indicator | Base line | A | B | C | D | E | F | G | H | I |
|---|---|---|---|---|---|---|---|---|---|---|
| Cumulative fuel-tax revenue, 2027–2050 (billion AUD) | 158 | 136 | 111 | 138 | 183 | 170 | 156 | 153 | 180 | 157 |
| of which RUC paid by RTs and ATs (billion AUD) | 49 | 39 | 34 | 39 | 73 | 65 | 48 | 46 | 70 | 48 |

*Notes: ZEHV = battery electric (BET) and hydrogen fuel cell (FCT) trucks. Parity is the year, rounded to the nearest year, in which the relative attractiveness of battery ATs, or of battery and fuel cell RTs combined, first equals that of the corresponding diesel trucks. Tailpipe $CO_2$ is calculated from diesel consumption at 2.7 kg $CO_2$ per litre. $CO_2$ intensity refers to the freight task of LCVs, RTs and ATs and is calculated at 2.7 kg $CO_2$ per litre of diesel and 2.3 kg per litre of petrol. Fuel-tax revenue is the fuel excise paid by LCVs plus the RUC paid by RTs and ATs, and excludes carbon-price revenue.*

## Supplementary File

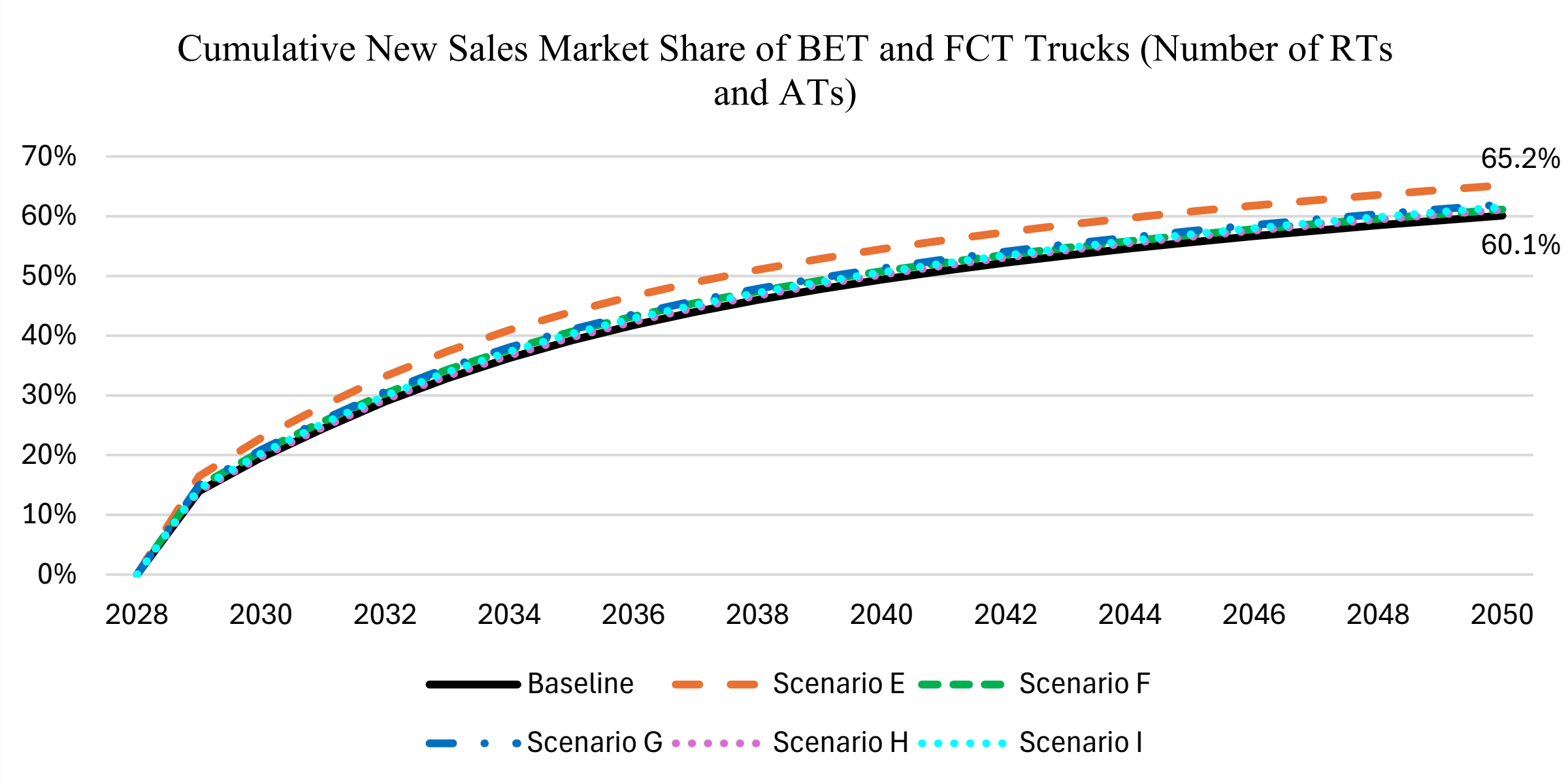

**Figure S1 | Cumulative share of BETs and FCTs in new RT and AT sales under policy scenarios E–I.**

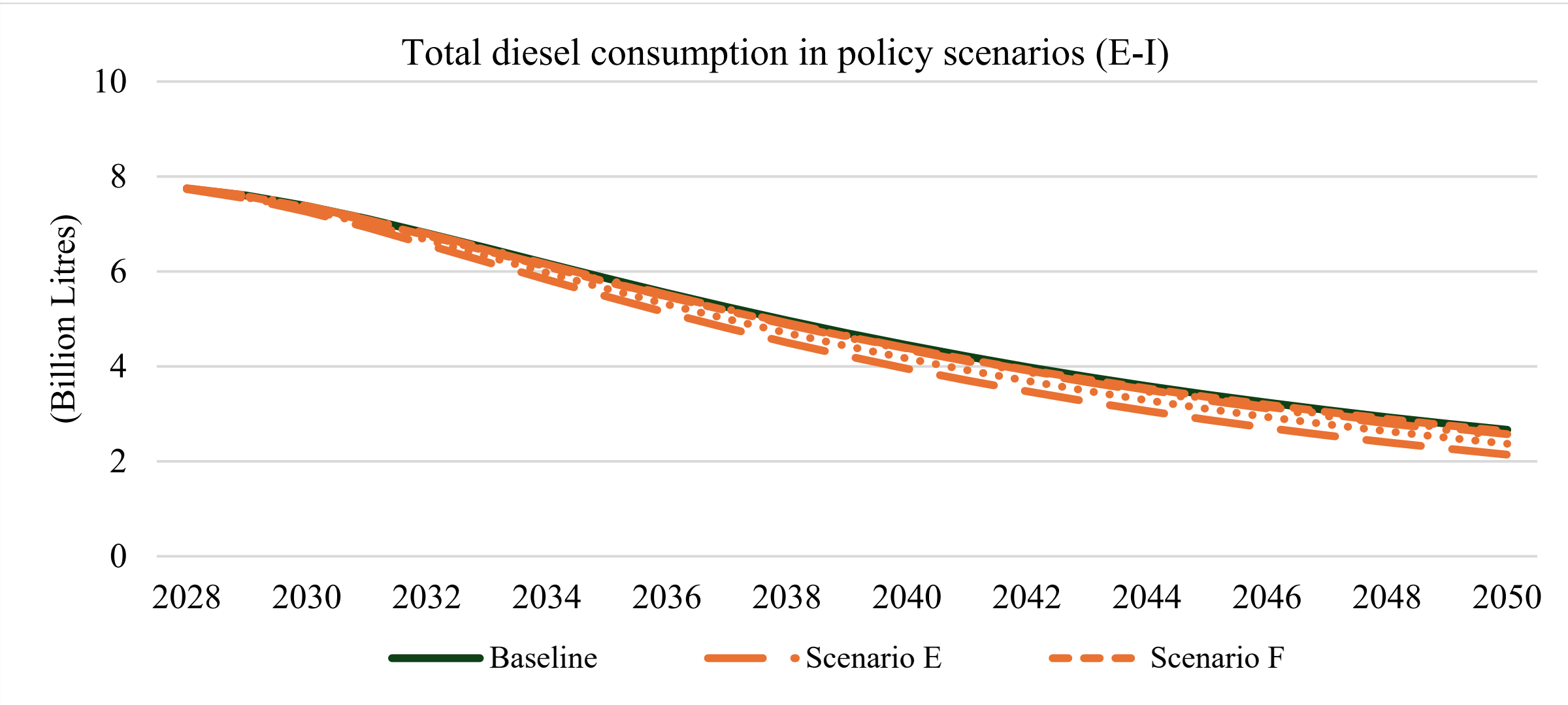

**Figure S2 | Diesel consumption of RTs and ATs (billion litres) in Scenarios E–I.**

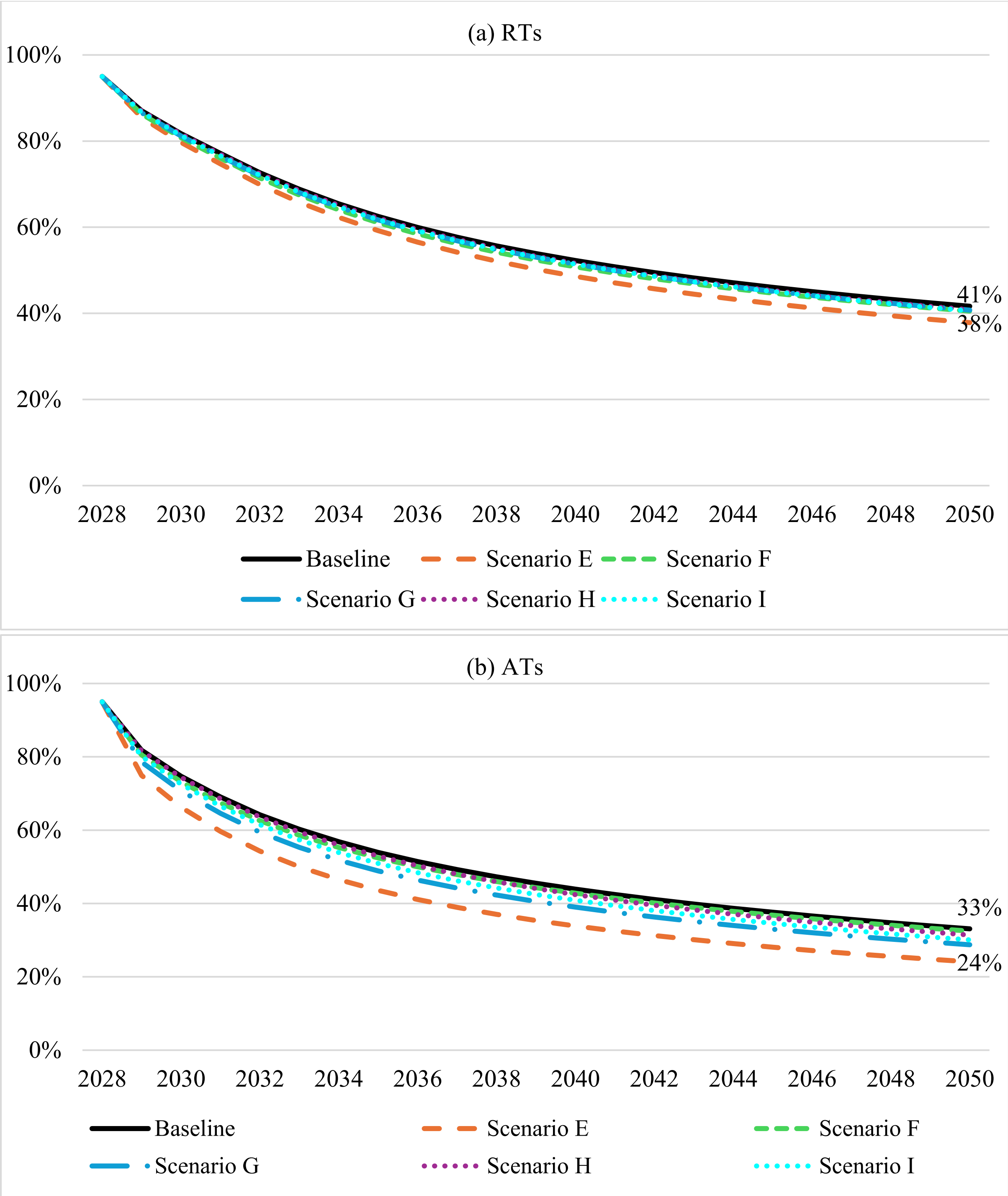

**Figure S3 | Cumulative share of diesel trucks in policy Scenarios (E-I): (a) RTs in new RT sales; (b) ATs in new AT sales.**